\documentclass[prb,amsfonts,amssymb,preprint,floats,superscriptaddress,aps]{revtex4-2}
	\usepackage{chngcntr}
	\usepackage{graphicx}
	\usepackage{color}
	\usepackage{amsmath}
	\usepackage{enumitem}
	\usepackage{amssymb}
	\usepackage{hyperref}
	\usepackage{cancel}
	\usepackage{ulem}
	\usepackage{multirow}

	\newcommand{\be}{\begin{equation}}
		\newcommand{\ee}{\end{equation}}
	
	\newcommand{\bea}{\begin{eqnarray}}
		\newcommand{\eea}{\end{eqnarray}}

	\newcommand{\la}{\left\langle}
	\newcommand{\ra}{\right\rangle}
	\newcommand{\lb}{\left[}
	\newcommand{\rb}{\right]}
	\newcommand{\lp}{\left(}
	\newcommand{\rp}{\right)}

	\newcommand{\K}{{\vec K}}
	\newcommand{\KK}{{\vec K}^{'}}
	\renewcommand{\vec}[1]{{\boldsymbol #1}}

\begin{document}
		\title{
		Superconductivity near spin and valley orders in graphene multilayers:
		a systematic study}
		\author{Zhiyu Dong}
		\author{Leonid Levitov}
		\affiliation{Department of Physics, Massachusetts Institute of Technology, Cambridge, MA 02139, USA}
		\author{Andrey V. Chubukov}
		\affiliation{W. I. Fine Theoretical Physics Institute, University of Minnesota, Minneapolis, MN 55455, USA}
		
		\begin{abstract}
		Spin excitations that soften near the onset of magnetic order have long been known to act as `paramagnon' pairing glue that can drive spin-triplet superconductivity. Recent findings of superconductivity in graphene bilayers and trilayers, occurring in the proximity of different itinerant ordered phases polarized in isospin (spin and valley),  have motivated us to conduct a comprehensive investigation of an isospin extension of the paramagnon pairing mechanism in the vicinity of spin/isospin orders. In each case, we identify a soft mode, associated with the order parameter fluctuations, that mediates pairing interaction. We develop an approach that relates the soft mode described through summation of the contributions most strongly divergent at the onset of spin/valley isospin orders. This interaction is not always attractive, but if it is,  it gives rise to an enhancement of superconducting $T_c$ in an appropriate pairing channel. In the cases when the pairing interaction is attractive, it leads to the formation of a superconducting state which can be either spin-triplet and valley-singlet or vice versa, depending on the specific isospin order type. These findings demonstrate that the occurrence of superconductivity in the vicinity of an itinerant magnetic phase is a generic phenomenon, closely mirroring experimental observations.
    			\end{abstract}
		\maketitle
		\newpage
		
		\section{Introduction}
		
		Recent experiments on graphene multilayers in a transverse electric field, such as Bernal bilayer graphene (BBG),
		revealed a complex phase diagram encompassing different spin and valley-ordered states with a cascade of phase transitions between them~\cite{de2021cascade,seiler2022quantum} and also found multiple superconducting phases near the onset of valley polarization~\cite{zhou2022isospin,zhang2022spin,holleis2023ising}.
Overall, the pattern of spin and valley polarized magnetic phases intertwined with superconductivity
bears some similarity to the magnetic and superconducting phases seen in twisted bilayer graphene \cite{cao2018unconventional,yankowitz2019tuning,lu2019superconductors,cao2021nematicity,cao2018correlated,
jiang2019charge,saito2021isospin,zondiner2020cascade,choi2021correlation,rozen2021entropic,pierce2021unconventional,
saito2021hofstadter}, a system that received considerable attention recently
(see \cite{Andrei2021} and references therein).
This triggered a renewal of interest in the low-energy physics of BBG, and also rhombohedral trilayer graphene (RTG), where a similar behavior has been reported~\cite{zhou2021superconductivity,zhou2021half} and analyzed~\cite{ghazaryan2021unconventional,chatterjee2021inter,You2022,chichinadze_22}.
		
		The electronic structure of BBG consists of two bands \cite{McCann_2013,McCann2006Landau}, with the wavefunctions supported by 
		the A sublattice in one 
		and the B sublattice in the other 
		layer. Without a transverse field, the two bands have quadratic dispersion and touch at the high-symmetry points ${\vec K}$ and ${\vec K}'$.  Applying a transverse electric field  opens up a band gap at charge neutrality \cite{McCann2006Landau,McCann2006Asymmetry} and changes the dispersion from quadratic to quartic.  While the bandwidth remains large, of order $1$eV, transverse field flattens the dispersion near the top of the valence band and the bottom of the conduction band. This gives rise to qualitative changes in the Fermi surface geometry and the ordered phasesas compared to those previously studied in
		unbiased bilayer graphene \cite{Nandkishore2010dynamical,Nandkishore2010quantum,Vafek2010,Jung2011lattice,macdonald2012pseudospin,Zhang2012Distinguishing,Cvetkovic2012,Throckmorton2014,Min2008,Nilsson2006}.
		
When the chemical potential $\mu$ is near the bottom of the conduction band (or the top of the valence band), filled states of fermions from one valley  are located  near ${\vec K}$ and from the other valley  near ${\vec K}'$. The Fermi surface near ${\vec K}$ (${\vec K}'$) does evolve with $\mu$ from three separate small pockets very near charge neutrality to a single Fermi surface  at somewhat larger doping. In between, the system passes through a van Hove singularity and features an annulus-type Fermi surface.
		
Flattened bands lead to a Stoner-type
 instability and several different spin- and valley-polarized phases. Several groups analyzed collective instabilities and superconductivity, specific to a particular geometry of the Fermi surface \cite{pantaleon2022superconductivity}. The superconducting phases seen in BBG and RTG share several common aspects of which the main one is that superconductivity occurs near the onset of one of spin/valley orders. Motivated by this, in this paper, we aim to understand the relation between spin
/valley orders and superconductivity
using a simple but broadly applicable framework that captures the essential aspects of BBG and RTG systems.
Namely, we
 employ
 a two-valley model that describes spin-$1/2$ fermions in valleys ${\vec K}$ and ${\vec K}'$ coupled by an electron-electron (e-e) repulsion interaction.  The intravalley interactions between fermions ($\vec K$ -- $\vec K$ or $\vec K'$ -- $\vec K'$) are taken to be different from the intervalley interactions ($\vec K$ -- $\vec K'$). 
 In particular, we assume that the interactions are insensitive to
the fine details of the electronic structure
in the valleys
(i.e., whether there is a single Fermi surface near each of the points $\vec K$ and $\vec K'$, or two annual Fermi surfaces, or three even smaller Fermi surfaces). We also account for exchange processes in which the interacting fermions undergo intervalley scattering, $\vec K\to\vec K'$ and $\vec K'\to\vec K$.
  
Our problem is simplified because several potentially viable effects happen to be small or absent in Bernal bilayer system. First, 
there is no  term with pair-hopping between the valleys as neither $2\K$ nor $2\KK$ is a reciprocal lattice vector.
 We can also legitimately ignore the Bloch wavefunction effects such as Berry phases and form factors in the e-e interaction.
  These effects are small
at realistic carrier densities, so long as the Fermi energy is much smaller than the bandgap induced by a transverse field \cite{dong2021isospin}.
		
Under these assumptions, the single-particle Hamiltonian reads:
		\be
H = \sum_{p,\alpha} \epsilon_p \psi_{1,p, \alpha}^\dagger \psi_{1,p,\alpha} +\sum_{p',\alpha} \epsilon_{p'} \psi_{2,p',\alpha}^\dagger \psi_{2,p',\alpha}
\label{1}
\ee
where the subscript $1,2$ denotes  valleys $\K$ and $\KK$, respectively.
The momenta $p$ and $p'$ label states in valleys $\K$ is near $\KK$. For a weak electron doping, filled states $\epsilon_p <\mu$ form small pockets near $\K$ and $\KK$.

The interaction Hamiltonian describing these low-energy fermions contains three main parts: $H_{\rm ee} = H_1 + H_2 + H_3$.  The first term is the density-density interaction between fermions in the same valley:
\be
H_{1} = \frac{U_1}2 \sum_{\alpha\beta;\gamma \delta} \left[\sum_{p k q}  \psi_{1,p+q, \alpha}^\dagger \psi_{1,k-q,\beta}^\dagger \psi_{1,k,\delta}\psi_{1,p, \gamma}+ \sum_{p' k' q} \psi_{2,p'+q, \alpha}^\dagger \psi_{2,k'-q,\beta}^\dagger \psi_{2,k',\delta}\psi_{2,p',\gamma}\right]  \delta_{\alpha \gamma} \delta_{\beta \delta}
\label{2}
\ee
where $\alpha$, $\beta$, $\gamma$, $\delta$ denote spin projections $\uparrow$ and $\downarrow$, the quantity $\delta_{\alpha\alpha'}$ is the Kronecker delta function. Here, the momenta ${\vec p}$, ${\vec K}$ are near $\K$, and ${\vec p}', {\vec K}'$ are near $\KK$, whereas the momentum transfer $\vec q$ is much smaller than $\vec Q=\K-\KK$ since typical $\vec q$ values are comparable to the Fermi momentum $p_F$ in each valley. The second term in the Hamiltonian $H_{\rm ee}$ is the interaction between fermion densities in different valleys
\be
H_{2} = U_2 \sum_{\alpha\beta;\gamma \delta} \sum_{p k' q} \psi_{1,p+q,\alpha}^\dagger \psi_{2,k'-q,\beta}^\dagger \psi_{2,k',\delta}\psi_{1,p,\gamma} \delta_{\alpha \gamma} \delta_{\beta \delta}
\label{3}
\ee
The third term 
involves simultaneous inter-valley scattering $\vec K\to \vec K'$, $\vec K'\to \vec K$,
\be
H_{3} = U_3 \sum_{\alpha\beta;\gamma \delta} \sum_{p k' q} \psi_{1,p+q,\alpha}^\dagger \psi_{2,k'-q,\beta}^\dagger \psi_{1,p,\delta}\psi_{2,k',\gamma} \delta_{\alpha \gamma} \delta_{\beta \delta}
.
\label{4}
\ee
The first two terms in the interaction Hamiltonian, $H_1$ and $H_2$, are the interactions with small momentum transfer. The couplings $U_1$ and $U_2$ are of the same order 
of magnitude though generally taking non-identical values. The 
interaction $H_3$ 
describes processes with momentum transfer near $\vec Q=\K-\KK$. The coupling strength $U_3$  is 
expected to be much smaller than $U_1$ and $U_2$ because dressed Coulomb interaction is substantially smaller at momentum ${\vec Q}$ than at momentum transfer of order $k_F$ within a given valley. For this reason, the inter-valley scattering
$U_3$
is often neglected. 
Here it will be retained because, as we will see, 
it lifts the degeneracy between several types of instabilities and is relevant for superconductivity.

The sign of $U_3$ is governed by the interplay of several effects. In a simple microscopic model, due to the exchange of two fermions in valleys ${\vec K}$ and ${\vec K}'$, $U_3$ takes a negative value, corresponding to the intervalley attraction. However, as shown in Ref.\cite{You2022}, the value of $U_3$ may become positive under the renormalization group flow in certain regimes of carrier density and transverse field. we therefore do not specify the sign of $U_3$ here, and keep the discussion generic.
		
		Within this model, there are four potential instabilities towards spin/valley orders, bilinear in fermions, two with zero momentum transfer, and two with momentum transfer ${\vec Q}= {\vec K}-{\vec K}'$. Of the two $\vec q=0$ orders, one is valley polarization - a valley imbalance type density order that makes Fermi energies in the two valleys unequal.
		The other is intra-valley ferromagnetism, with either ferromagnetic or antiferromagnetic ordering between valleys,
		each
		with a three-component order parameter due to the global spin rotation $SU(2)$ symmetry that acts on electrons in both valleys. Of the two orders with the momentum transfer ${\vec Q}$, one is charge-density-wave (CDW) with a two-component complex order parameter due to the valley conservation $U(1)_v$ symmetry. 
		 The other is spin-density-wave (SDW) with a $3\times 2=6$-component complex order parameter due to $U(1)_v \times SU(2)$ symmetry. We note that, the CDW and SDW orders are also known as spin-singlet and spin-triplet inter-valley coherence (IVC) orders, respectively\cite{chatterjee2022inter}. The total number of the order parameter components is 15 (one for valley polarization, $3+3$ for intra-valley ferromagnetism, $2$ for a complex CDW order, and $6$ for a complex SDW order).
		
		When inter-valley scattering is negligible and density-density interactions for fermions within a given valley and in different valleys are identical, the conditions for all  four instabilities are identical. In this case,  15 order parameters form the adjoint representation of the SU(4) symmetry group~\cite{chichinadze_prl}.  The ordered state can be a single order, one of the four,  or a combination of different orders, e.g. a valley polarization order and ferromagnetism within just one valley~\cite{chichinadze_prl,chichinadze_22}.  Here we differentiate between density-density interactions with a valley and between valleys and include inter-valley scattering (which as we show, is relevant to superconductivity). In this situation, different orders emerge at different temperatures/doping levels, and one can consider separately a given order and superconductivity near it.
		
		In a recent paper~\cite{dong_23} we analyzed the instability towards valley polarization and superconductivity near its onset. In this work, 
		the results of 
		the analysis of three other potential orders in BBG  and superconductivity near each of these orders are reported. We specifically address three main questions: 
		\begin{enumerate}
		\item[(i)] Can the  pairing interaction be enhanced near the onset of a given order? 
		\item[(ii)] Can the pairing interaction be viewed as mediated by soft fluctuations of the corresponding  density or spin order parameter?
		\item[(iii)] Can the pairing interaction 
		become attractive in at least one of the channels?
		\end{enumerate}
		It is known that for systems without an additional valley degree of freedom, the answer to all three questions is generally in the affirmative (see Sec.\ref{sec: SU(2)}	). 
		In the present case of Bernal bilayer graphene the situation is more involved since Cooper pairs with zero total momentum comprise fermions in different valleys, $\vec K$ and $\vec K'=-\vec K$.
		Without intra-valley scattering, the pairing vertex for such fermions does not have a component coming from anti-symmetrization (fermionic exchange) and hence has no component that would couple to soft spin fluctuations near a magnetic transition.  Near the onset of density order, the valley exchange interaction is not relevant. This constitutes a new problem that has not been analyzed in previous literature. 
		
		In Ref.~\cite{dong_23} we 
		identified a pairing interaction 
		mediated by soft density fluctuations associated with valley-polarization instability.  
		While the sign of this interaction turns out to be repulsive, we argued in~\cite{dong_23} that superconductivity is still possible if an additional in-plane magnetic field is applied which couples to electron spin by Zeeman interaction. A similar pairing mechanism was found to be operational in the presence of an Ising spin-orbit interaction
		which effectively induces a valley-odd Zeeman field.  Superconductivity at a finite field is consistent with the data in~\cite{zhou2021BLG}, and the one at a finite spin-orbit coupling is consistent with the data in \cite{zhang2022spin} for BBG placed on top of a  monolayer of tungsten diselenide  as by all accounts WSe$_2$ induces spin-orbit coupling.
		
		Here we consider the problem near the onset of intra-valley ferromagnetism, wherein the pairing interaction is again enhanced
 and
  can be viewed as mediated by spin fluctuations. We identify pairing channels in which attraction arises in a more robust manner than in Ref.\cite{dong_23}. There are two types of intra-valley ferromagnetism: an inter-valley ferromagnetism, a phase in which the spin polarizations in two valleys are parallel, and an inter-valley antiferromagnetism, a phase in which spin polarizations in the two valleys are antiparallel. 
    Near the onset of 
    inter-valley ferromagnetism, spin-mediated interaction is attractive in spin-triplet, valley singlet channel.  Near the onset of 
     inter-valley antiferromagnetism, the attractive interaction is in spin-singlet, valley-triplet channel.  While this result agrees with what one could expect on general grounds, we emphasize that this holds only when we include $U_3$ scattering. We show that to obtain such spin-mediated attraction, 
     one has to include diagrammatic series to all orders in $U_3$.
     Near CDW and SDW transitions, the pairing interaction can be viewed as mediated by density and spin fluctuations, respectively.  Yet, the pairing interaction near an SDW instability is repulsive in both singlet and triplet channels and does not give rise to superconductivity, at least at zero fields and without spin-orbit coupling. The one near CDW instability is repulsive in the spin-triplet channel, but attractive in the spin-singlet channel.  We emphasize that this holds for repulsive density-density interaction and an arbitrary sign of inter-valley scattering.

\section{An SU(2) spin-$1/2$ model}	
\label{sec: SU(2)}	
			Before addressing the full two-valley problem, Eqs.\eqref{1}--\eqref{4}, here we consider a simpler problem -- a one-valley problem described by an SU(2) spin-$1/2$ model. We will work out a solution to this SU(2) problem as an illustration for our approach.
			Later we will use the same approach to study an $SU(4)$ problem that encompasses spin and valley degrees of freedom, which is the main focus of this paper.

The single-particle Hamiltonian in this one-valley model is simply 	$H_0^{\rm{SU(2)}} = \sum_{p,\alpha} \epsilon_p \psi_{p, \alpha}^\dagger \psi_{p,\alpha} $, and the interaction between electrons can be modeled using a short-range repulsion:
		\be
		H^{\rm{SU(2)}}_{\rm int} = \frac{U}2 \sum_{\alpha\beta;\gamma \delta} \sum_{p k' q} \psi_{p+q,\alpha}^\dagger \psi_{k'-q,\beta}^\dagger \psi_{k',\delta}\psi_{p,\gamma} \delta_{\alpha \gamma} \delta_{\beta \delta}
	\ee
	where $\alpha,\beta,\gamma,\delta$ are spin indices.

 We consider the pairing interaction arising from this Hamiltonian.
 A generic pairing interaction between spin $1/2$ fermions is
 	\begin{align}
 \label{n_1}
	\Gamma_{\alpha\beta;\gamma\delta}(k,-k;p,-p)& = U^{\rm{SU(2)}}_a\delta_{\alpha\gamma}\delta_{\beta\delta} - U^{\rm{SU(2)}}_b \delta_{\alpha\delta}\delta_{\beta\gamma}
	\nonumber \\ & =  \lp U^{\rm{SU(2)}}_a- \frac{U^{\rm{SU(2)}}_b}{2}\rp \delta_{\alpha\gamma}\delta_{\beta\delta}- \frac{U^{\rm{SU(2)}}_b}{2} \vec \sigma_{\alpha\gamma}\cdot \vec \sigma_{\beta\delta}
\end{align}
 where $U^{\rm{SU(2)}}_a$ is the fully dressed irreducible interaction with momentum transfer $k-p$ and  $U^{\rm{SU(2)}}_b$ is the fully dressed interaction with momentum transfer $k+p$.  Each dressed interaction is given by infinite series of diagrams in which Kohn-Luttinger terms are the leading ones.
  In the last line in (\ref{n_1}) we used the Fierz identity $\delta_{\alpha\delta}\delta_{\beta\gamma} = (1/2) (\delta_{\alpha\gamma}\delta_{\beta\delta} +\vec \sigma_{\alpha\gamma}\cdot \vec \sigma_{\beta\delta})$.

\begin{figure}
			\centering
			\includegraphics[width=0.55\textwidth]{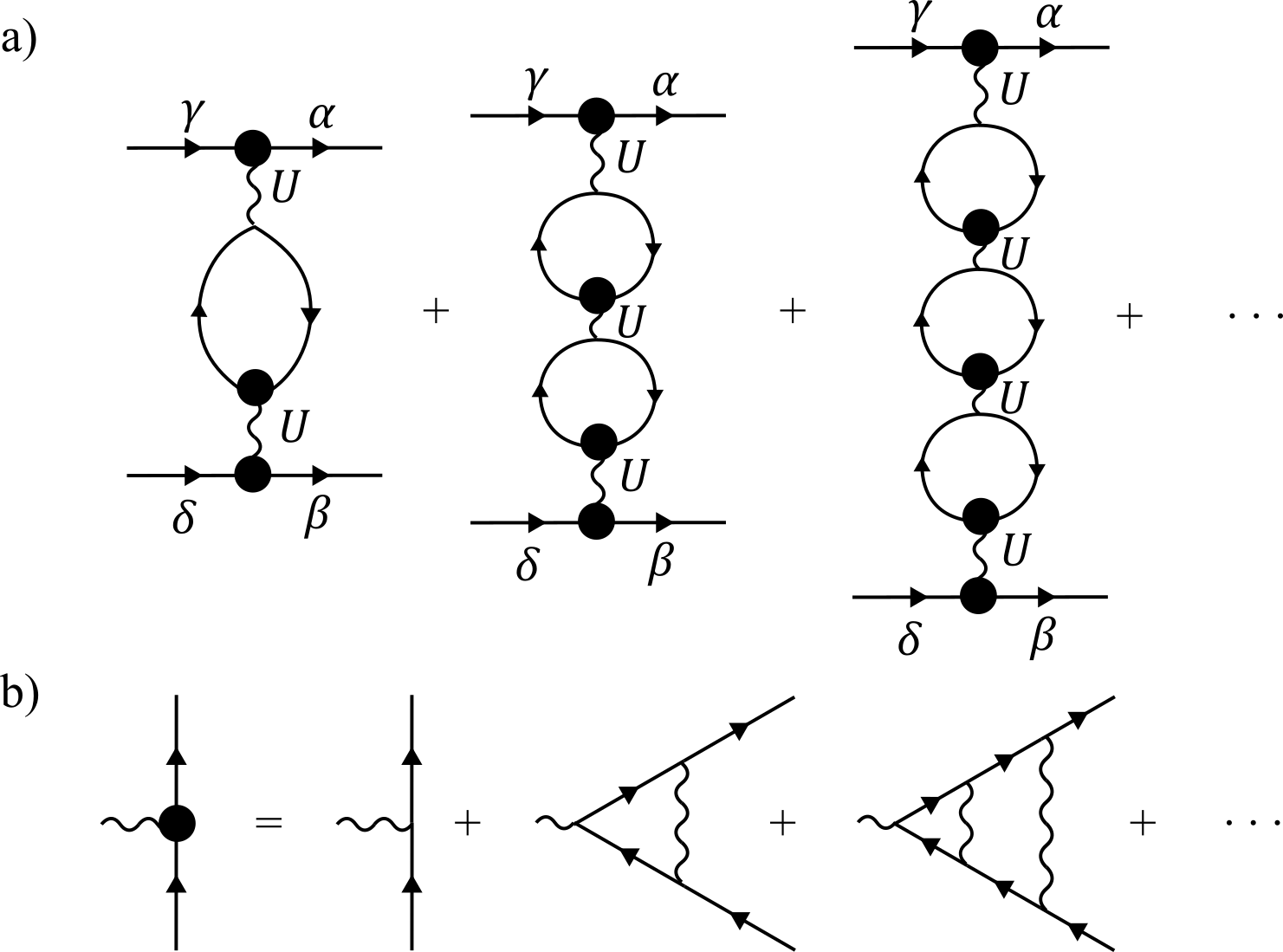}
			\includegraphics[width=0.35\textwidth]{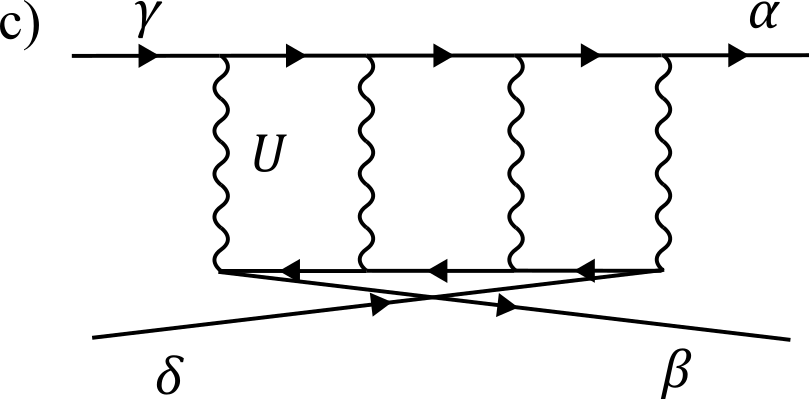}
			\caption{Diagrammatic representation for the leading contributions to
 the pairing vertex (the antisymmetrized pairing interaction) with zero momentum transfer $\Gamma_{\alpha \beta; \gamma \delta} (k, -k,k,-k)
 = U^{\rm{SU(2)}}_a\delta_{\alpha\gamma}\delta_{\beta\delta}- U^{\rm{SU(2)}}_b \delta_{\alpha\delta}
 \delta_{\beta\gamma}$.
 This pairing vertex
  is obtained from the diagrams most strongly divergent near the Stoner instability in the SU(2) spin-$1/2$ model.
  Panels (a) and (b) --
 diagrams for the direct component  $U^{\rm{SU(2)}}_a$ (interaction with momenta $(k,-k;k,-k)$).
 Panel (c) -- diagrams for the component  $U^{\rm{SU(2)}}_b$ with the two outgoing fermions interchanged (interaction with momenta $(k,-k;-k,k)$).
 See Eqs.(\ref{eq:Ua}) and (\ref{eq:Ub}), and accompanying text.}
			 \label{fig:instability_SU2}
		\end{figure}

  Our goal is to obtain the pairing interaction near an instability in the particle-hole channel.
 It is well-known that in a one-valley system with a repulsive interaction, a potential instability is
 a Stoner-like one, towards a magnetic order.
For simplicity, we consider a ferromagnetic instability. It occurs when
 $U \Pi (0) =1$, where $\Pi = \Pi (0)$ is the static polarization function at vanishing momentum.
Our goal then is to identify series of diagrams for $U^{\rm{SU(2)}}_a$ and $U^{\rm{SU(2)}}_b$,
   which contain the factor $1/(1- U \Pi (0))$.  Accordingly, we consider pairing vertex $\Gamma_{\alpha\beta;\gamma\delta}(k,-k;p,-p)$ with $p \approx k$. A simple experimentation shows that for $U^{\rm{SU(2)}}_a$ (the dressed interaction that scatters
a pair from momenta $(k,-k)$ to momenta $(k,-k)$  with zero momentum transfer),
 the relevant diagrammatic series are the ones shown
  in Fig.\ref{fig:instability_SU2}
a-b. 		
	These series yield
	\be \label{eq:Ua}
	U^{\rm{SU(2)}}_a = \lp \frac{1}{1-U\Pi (0)}\rp ^2 \frac{U}{1+\frac{2U\Pi}{1-U\Pi (0)}} = \frac{U}{1-U^2\Pi^2(0)}
	\ee
	The relevant diagrammatic series for the dressed $U^{\rm{SU(2)}}_b$ (the interaction that scatters a pair from momenta $(k,-k)$ to momenta $(-k,k)$ with a momentum transfer of $2k$), are the ones shown in Fig.\ref{fig:instability_SU2}
c.
 These series yield
	\be \label{eq:Ub}
	U^{\rm{SU(2)}}_b = \frac{U}{1-U\Pi (0)}
	\ee
 Adding these contributions, we obtain the effective pairing interaction in the form
	\begin{align}
	\Gamma_{\alpha\beta;\gamma\delta}(k,-k;k,-k)& =
 \frac{U}{2} \lp \frac{\delta_{\alpha\gamma}\delta_{\beta\delta}}{1 + U \Pi (0)} - \frac{\vec \sigma_{\alpha\gamma}\cdot \vec \sigma_{\beta\delta} }{1 - U \Pi (0)} \rp
	\end{align}
The divergence of pairing interaction near a FM instability arises from the second term, which
 can be viewed as the pairing interaction, mediated by soft ferromagnetic spin fluctuations~\cite{Scalapino2012}.
Such an interaction is attractive in spin-triplet channel and leads to
spin-triplet pairing.
 A similar analysis near an antiferromagnetic instability yields a pairing interaction mediated by soft antiferromagnetic fluctuations. Such an  interaction is attractive in the spin-singlet channel~(see e.g., \cite{Monthoux2007}).


		\section{The two-valley model}
		
		Below we apply the diagrammatic approach introduced above to study the two-valley
 model
 defined in Eqs.\eqref{1}--\eqref{4}.  As we stated in the Introduction, we neglect momentum variations of the interactions, i.e., treat $U_i$ as constants.
Along the same lines, we treat the polarization bubbles $\Pi(q)$ with small q, both for fermions in the same valley and in different valleys,
as some positive $\Pi (0)$,
 and treat  the polarization bubble $\Pi(q)$ with $q \approx {\bf Q} ={\bf K} - {\bf K}'$ as some positive $\Pi (Q)$.
We assume that  $\Pi (0)$ and $\Pi(Q)$ are enhanced due to large density of states in flattened bands, but
 neglect the fine structure of $\Pi (q)$ which occurs when (i) the Fermi sea in each isospin consists of more than one pocket or (ii) the Fermi pocket is single-piece but non-circular.
 Because low-energy fermionic states exists only near ${\bf K}$ and ${\bf K}'$ and hence the momenta $k$ and $p$ in (\ref{n_1})  are constrained to the vicinity of these points, this last  restriction implies that the
  pairing gap $\Delta (k) = \Delta ({\bf K})$  is a constant.
  Without the isospin variable, this would restrict the pairing to only the spin-singlet channel. In our
   case, however,
   both spin-singlet and spin-triplet pairing are possible because a fermionic pair with momenta ${\vec K}$ and $-{\vec K}$ is made of fermions from different valleys, and the gap function can be either valley-symmetric or valley-antisymmetric. A generic requirement that a superconducting order parameter must change sign upon the exchange of the two fermions  then implies that the valley-symmetric gap function is spin-singlet and the valley-antisymmetric gap function is spin-triplet.
  The valley-antisymmetric gap changes sign upon replacement ${\bf k}$ to $-{\bf k}$ and in this respect is analogous to a gap function near a FM instability in a one-valley system.
		
		\section{Ordered states}
		In this section, we analyze the instability conditions for different ordered states with order parameters, bilinear in fermions.  Simple experimentation shows that the order parameters,  which  one can create out of  $\psi_1$ and $\psi_2$, include  \begin{enumerate}
			\item[a)]
			Valley polarization: an order that splits fermionic densities in the two valleys
			\be
			\Delta_{\rm VP} = \frac{2}{N} \sum_{\alpha \beta} \left[\sum_{p}
			\la \psi^\dagger_{1,p, \alpha} \delta_{\alpha \beta} \psi_{1,p,\beta}\ra -  \sum_{p'} \la\psi^\dagger_{2,p',\alpha} \delta_{\alpha \beta}\psi_{2,p',\beta}\ra \right]
			\label{5}
			\ee
			This is a single scalar order parameter.
			\item[b)]
			Intra-valley ferromagnetism
			\be
			{\vec \Delta}_{\rm 1, FM} = \frac{2}{N} \sum_{\alpha \beta} \sum_p \la \psi^\dagger_{1,p,\alpha} {\vec \sigma}_{\alpha \beta} \psi_{1,p,\beta}\ra,  ~~{\vec \Delta}_{\rm 2,FM} = \frac{2}{N} \sum_{\alpha \beta} \sum_{p'} \la \psi^\dagger_{2,p',\alpha} {\vec \sigma}_{\alpha \beta} \psi_{2,p',\beta}\ra ,
			\label{6}
			\ee
			Each order parameter is a 3-component vector, so the total number of order parameter components here is six.
			
			When  ${\vec \Delta}_{\rm 1,FM}$ and ${\vec \Delta}_{\rm 2,FM}$ are parallel to each other, the ordered state is an inter-valley 
			ferromagnet,  when the two are antiparallel to each other the order is ferromagnetic within the valley and antiferromagnetic between the valleys.
			\item[c)]
			A CDW order with momentum ${\vec Q} = \K -\KK$
			\be
			\Delta_{\rm CDW} = \frac{2}{N} \sum_{\alpha \beta} \sum_{p,p'} \la \psi^\dagger_{1,p, \alpha} \delta_{\alpha \beta} \psi_{2,p',\beta}\ra \delta_{{\bf p} - {\bf p'} - {\vec Q}}
			\label{7}
			\ee
			this order parameter is a complex function, i.e., $\Delta_{\rm CDW}$ and $\Delta^*_{\rm CDW}$ are not identical. Both $\Delta_{\rm CDW}$ and $\Delta^*_{\rm CDW}$ are scalar order parameters, so the total number of order parameter components is two.
			\item[d)]
			A SDW order with momentum ${\vec Q} = \K -\KK$
			\be
			{\vec \Delta}_{\rm SDW} = \frac{2}{N} \sum_{\alpha \beta} \sum_{p,p'} \la \psi^\dagger_{1,p,\alpha} {\vec \sigma}_{\alpha \beta} \psi_{2,p',\beta}\ra  \delta_{{\bf p} - {\bf p'} - {\vec Q}}
			\label{8}
			\ee
			This is again a complex order parameter:  ${\vec \Delta}_{\rm SDW}$ and ${\vec \Delta}^*_{\rm SDW}$ are not identical. Each order parameter has three spin components. Hence the  total number of components is six.
		\end{enumerate}
		
		Altogether, there are 15 order parameter components, some of which are degenerate. To analyze when each order develops spontaneously, we do a standard analysis~\cite{chichinadze_20,*chichinadze_20a}: introduce the  trial order parameter and construct a self-consistent equation on it by collecting ladder and bubble renormalizations  with transferred momentum either zero or ${\vec Q}$. We show self-consistent equations diagrammatically in Fig. 1. Each self-consistent equation has a non-zero solution at the onset of the corresponding order.  Alternatively, one could introduce an infinitesimally small trial vertex, renormalize it by inserting ladder and bubble diagrams with the corresponding momentum transfer, and obtain the susceptibility.  This susceptibility diverges at the onset of spontaneous order.
		
					\begin{figure}
			\includegraphics[width=0.99\textwidth]{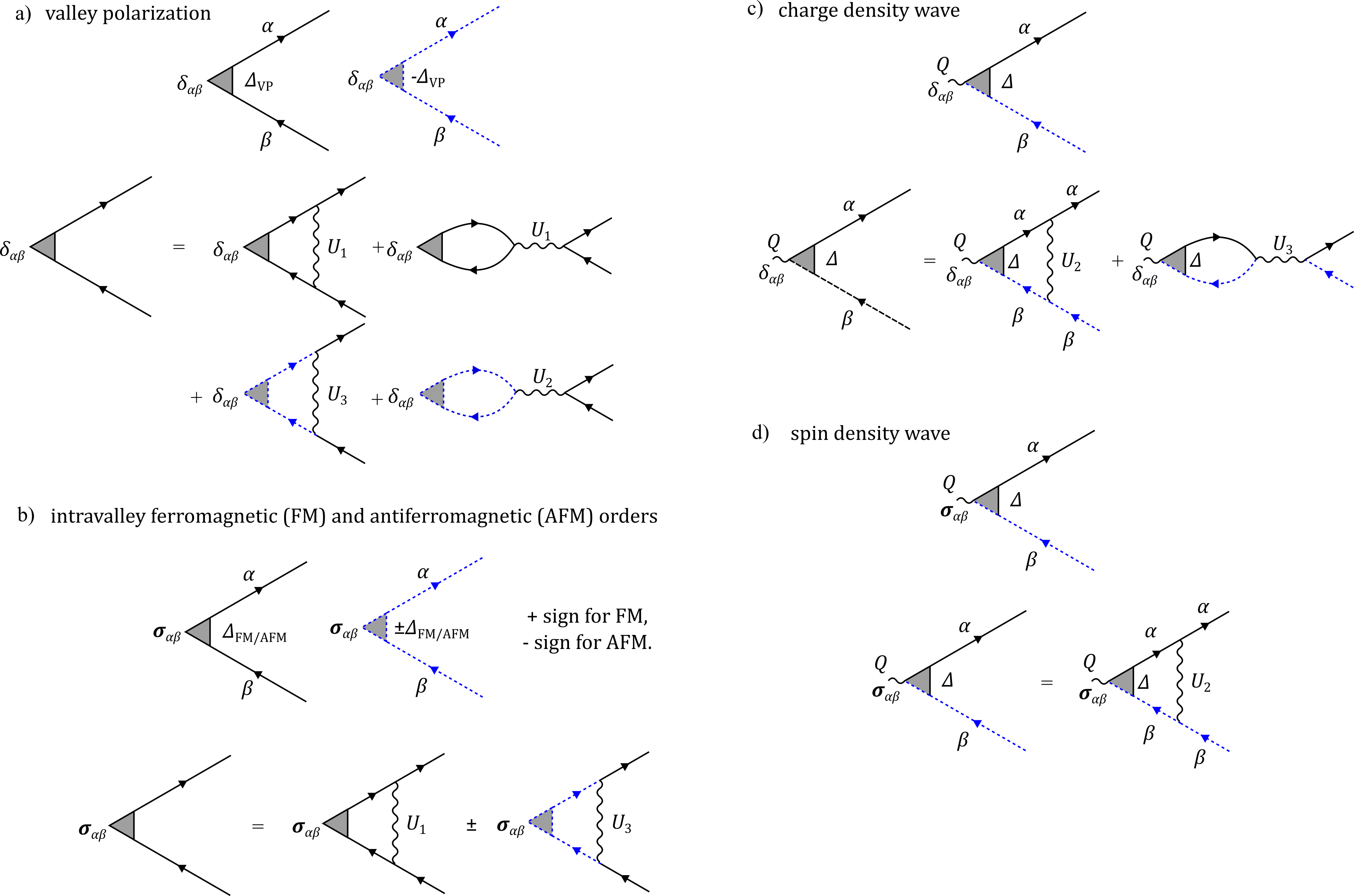}
			\centering
			\caption{Diagrammatic representation of the order parameters for different order types arising within the model given in Eqs.\eqref{1}--\eqref{4}. Solid and dashed lines colored in black and blue represent fermions in valleys $\vec K$ and $\vec K'$, respectively. Panels (a), (b), (c) and (d) detail the Stoner mean field description for valley-polarization order, ferromagnetic order, CDW order and SDW order, respectively. The CDW and SDW orders are also known in the literature as the spin singlet and spin triplet inter-valley coherence orders, respectively.
In each panel the diagrams illustrate the mean-field self-consistency relations described in the text, see Eqs. \eqref{9}, \eqref{10+}, \eqref{10-}, \eqref{11} and \eqref{12}.
			  }\label{fig:instability_abcd}
		\end{figure}
		
		Solving the diagrammatic equations we find that
		\begin{enumerate}
			\item[a)]
			Valley polarization instability emerges when
			\be
			(2U_2-U_1-U_3)
\Pi (0)=1,
			\label{9}
			\ee
			where
  we remind that $\Pi (0)$ is
the particle-hole polarization bubble at zero momentum transfer.
Like we said, a
sufficient information for our purposes is that 
$\Pi (0)$ is positive and is enhanced due to flat dispersion $\epsilon_p$ near $\K$ and $\KK$.
			\item[b)]
			To detect the onset of intra-valley ferromagnetism, leading to either ferromagnetic or antiferromagnetic
			ordering of ${\vec \Delta}_{\rm 1,FM}$ and ${\vec \Delta}_{\rm 2,FM}$, we need to solve the set of two coupled equations for these order parameters. The coupling is via $U_3$.  The results are: an inter-valley ferromagnetic order develops at
			\be
			(U_1 +U_3) \Pi (0)  =1,
			\label{10+}
			\ee
			and an inter-valley antiferromagnetic order with ferromagnetism within a valley develops at
			\be
			(U_1 -U_3) \Pi (0) =1,
			\label{10-}
			\ee
			These two orders are described by the order parameters $\Delta_+ = \Delta_{\rm 1,FM} + \Delta_{\rm 2,FM} $ and $\Delta_- = \Delta_{\rm 1,FM} - \Delta_{\rm 2,FM} $, respectively. We call the former order FM$^+$ and the latter  FM$^-$. For $U_3 >0$,  the instability towards inter-valley ferromagnetism develops first, and for $U_3 <0$, the instability towards inter-valley antiferromagnetism develops before the one towards inter-valley ferromagnetism.
			\item[c)]
			A CDW instability develops at
			\be
			(U_2 -2 U_3) \Pi (Q) =1,
			\label{11}
			\ee
			where, we remind,
 $\Pi (Q)$
 is the particle-hole polarization bubble at momentum transfer ${\vec Q} = \K - \KK$. It is comparable in magnitude but not identical to $\Pi (0)$ (see e.g., Ref. \cite{You_21}).
			\item[d)]
			An SDW instability develops at
			\be
			U_2\Pi (Q) =1,
			\label{12}
			\ee
			For $U_3 >0$,  SDW develops before CDW, and for $U_3 <0$  CDW order develops first.
		\end{enumerate}
		If $U_3$ and the difference between $U_1$ and $U_2$ and between $\Pi (0)$ and $\Pi (Q)$ were negligibly small (i.e., $\Pi (0) \approx\Pi (Q) \approx \Pi$  , all instabilities would occur at the same $U_1 \Pi =1$. The fifteen order parameters would then form the adjoint representation of the SU(4) symmetry group~\cite{chichinadze_prl}.  In our analysis, we keep $U_1$ and $U_2$ different, $U_3$ finite, and $\Pi (Q) \neq \Pi (0)$.   In this situation, each order develops separately, and one can find the condition at which a given order develops before the others.
		
		\begin{figure}
			\includegraphics[width=0.5\textwidth]{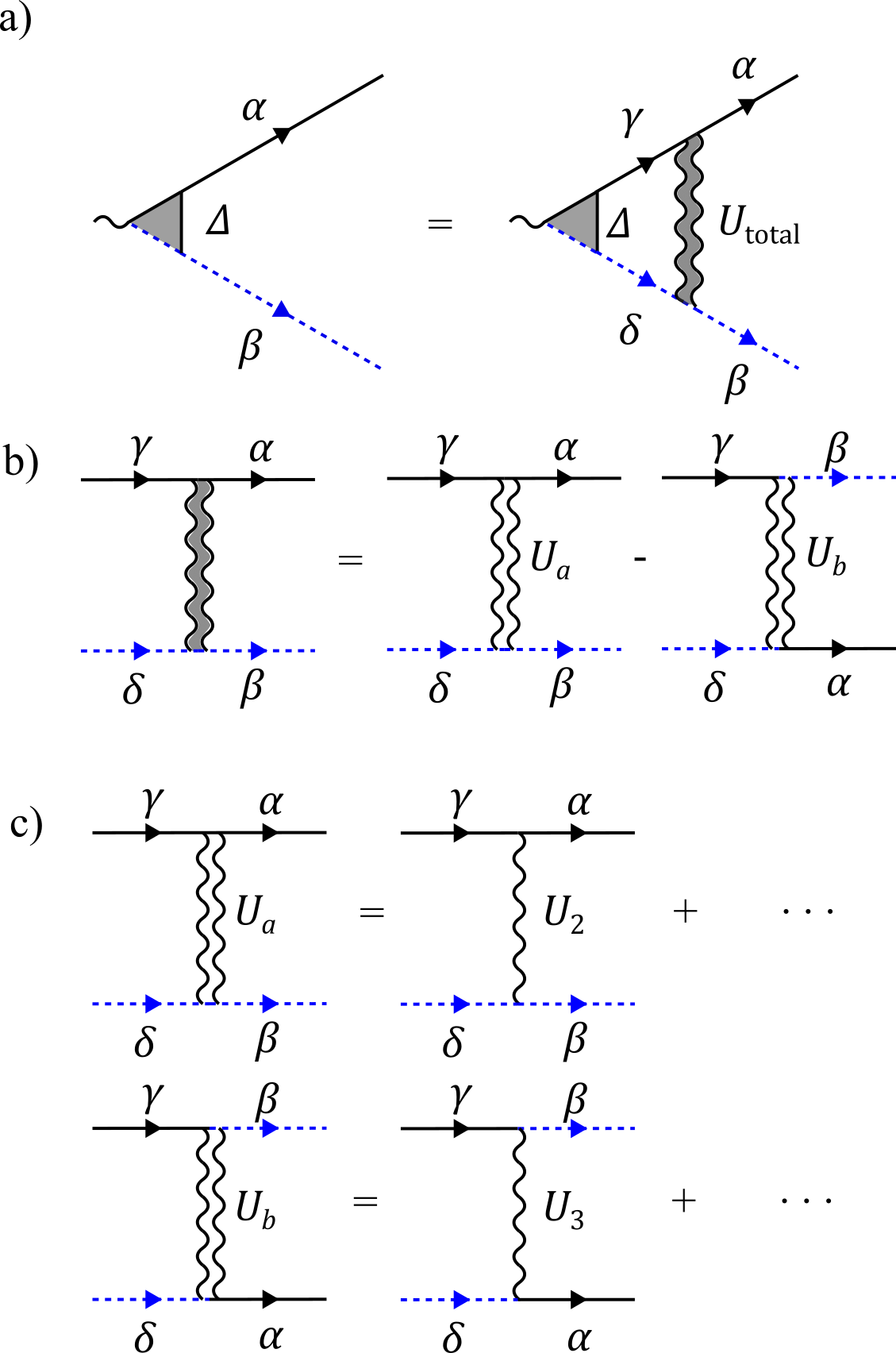}
			\centering
			\caption{a) Pairing interaction between carriers in valleys $\vec K$ and $\vec K'$, represented by solid and dashed lines colored in black and blue, respectively. Shown is a diagrammatic representation of a selfconsistency relation for the superconducting order parameter. The shaded wavy line labeled $U_{\rm total}$ is a properly antisymmetrized pairing interaction detailed in panels (b) and (c). The paring interactions shown in b) by open wavy lines accounts for the intravalley and intervalley scattering processes, with the minus sign arising due to anticommutation of fermions in the states $\vec K\alpha$ and $\vec K'\beta$. 
			The interactions $U_a$ and $U_b$ are the renormalized intervalley density-density interaction and the intervalley exchange interaction, as detailed in panel (c).
			}\label{fig:pairing_interaction}
		\end{figure}
				
		\section{Pairing interaction}

		In this section, we analyze the pairing interaction near the onset of each of the potential orders. As stated above, the three key questions of interest are: (i) whether the pairing interaction is enhanced, (ii) whether it can be viewed as mediated by soft fluctuations of the order parameter, either in
 the
 density or in the spin channel, and (iii) whether the pairing interaction is attractive.
		
		To analyze the pairing, we
introduce a trial order parameter in the particle-particle channel, 
$\Delta$
, and obtain a self-consistent equation on this order parameter.  The self-consistent equation is shown diagrammatically in Fig.
   3a.
   The pairing interaction has the same form as in Eq. (\ref{n_1}), but now the two components are
   $U^{\rm{SU(4)}}_a$ and $U^{\rm{SU(4)}}_b$. To simplify notations, below we label them as just $U_a$ and $U_b$.
    We have
%
		\be
		\Gamma_{\alpha \beta; \gamma \delta} (k,-k;k,-k)
		 = U_a \delta_{\alpha \gamma} \delta_{\beta \delta} - U_b \delta_{\alpha \delta} \delta_{\beta \gamma}
		\label{14}
		\ee
		Using the Fierz identity, as before,  this can be equivalently expressed as
		\be
		\Gamma_{\alpha \beta; \gamma \delta} (k,-k;k,-k)
		= U_d \delta_{\alpha \gamma} \delta_{\beta \delta} + U_s {\vec \sigma}_{\alpha \delta} \cdot {\vec \sigma}_{\beta \gamma}
		\label{15}
		\ee
Here		
 $U_d = U_a- U_b/2$ and $U_s = - U_b/2$ are density and spin components of the
  pairing interaction.
   At the bare level,
   $U_a = U_2$ and $U_b = U_3$ (see Fig. \ref{fig:pairing_interaction}c).  The fully dressed $U_a$ and $U_b$ include the renormalizations from insertions of particle-hole bubbles.

		\subsection{Valley polarization}
		
		\begin{figure}
		\includegraphics[width=0.8\textwidth]{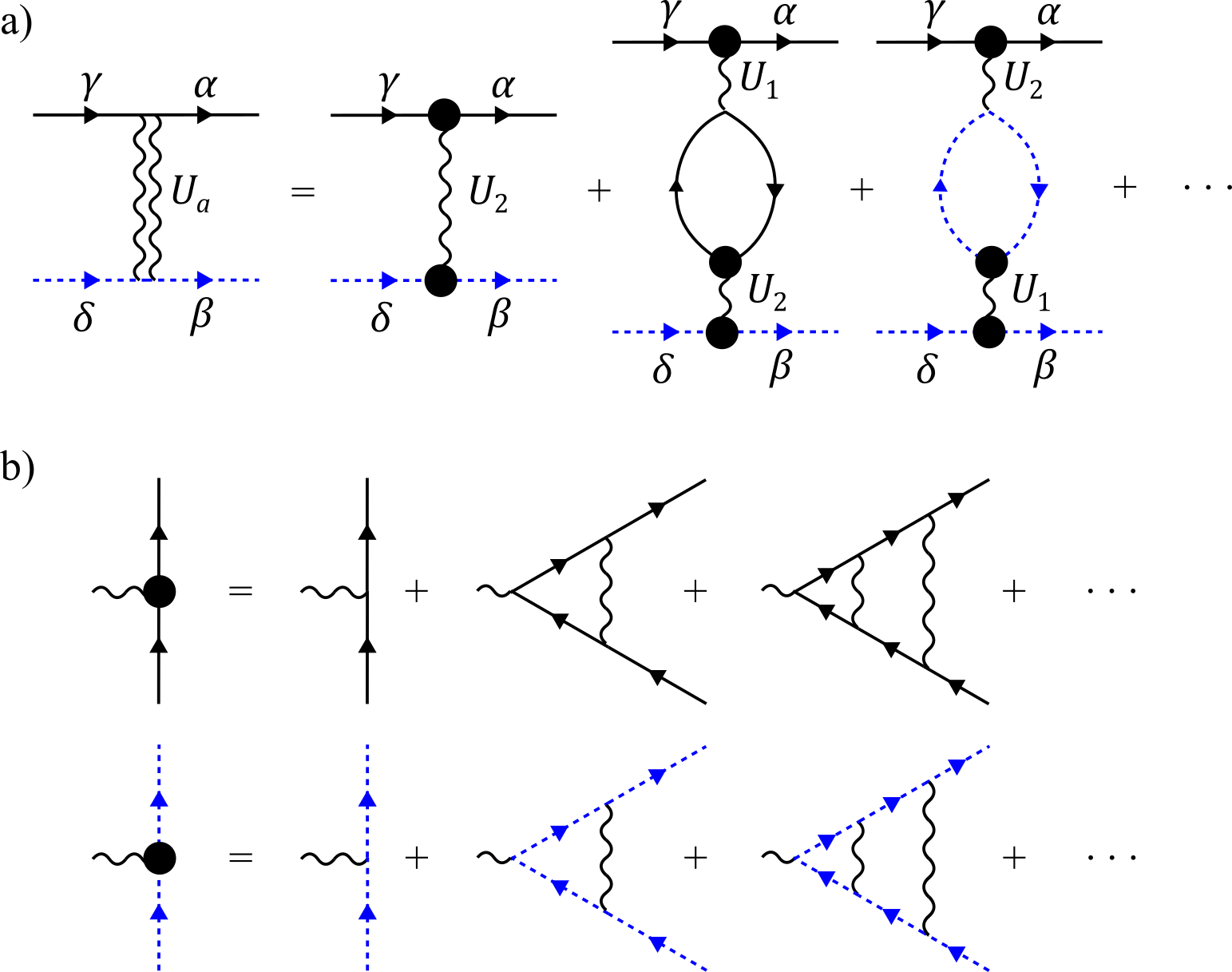}
		\centering
		\caption{Panel (a) -- diagrammatic expression for the effective interaction $U_a$ when $U_3$ is set to zero.
 We keep only the diagrams which contain $\Pi (0)$.
The momenta along the upper line are approximately $(k,k)$ and the ones along the lower line are $(-k,-k)$.
The dressed vertices (the black dots) are given by the diagrams in panel (b)
}\label{fig:Ua}
		\end{figure}
		\begin{figure}
			\includegraphics[width=0.6\textwidth]{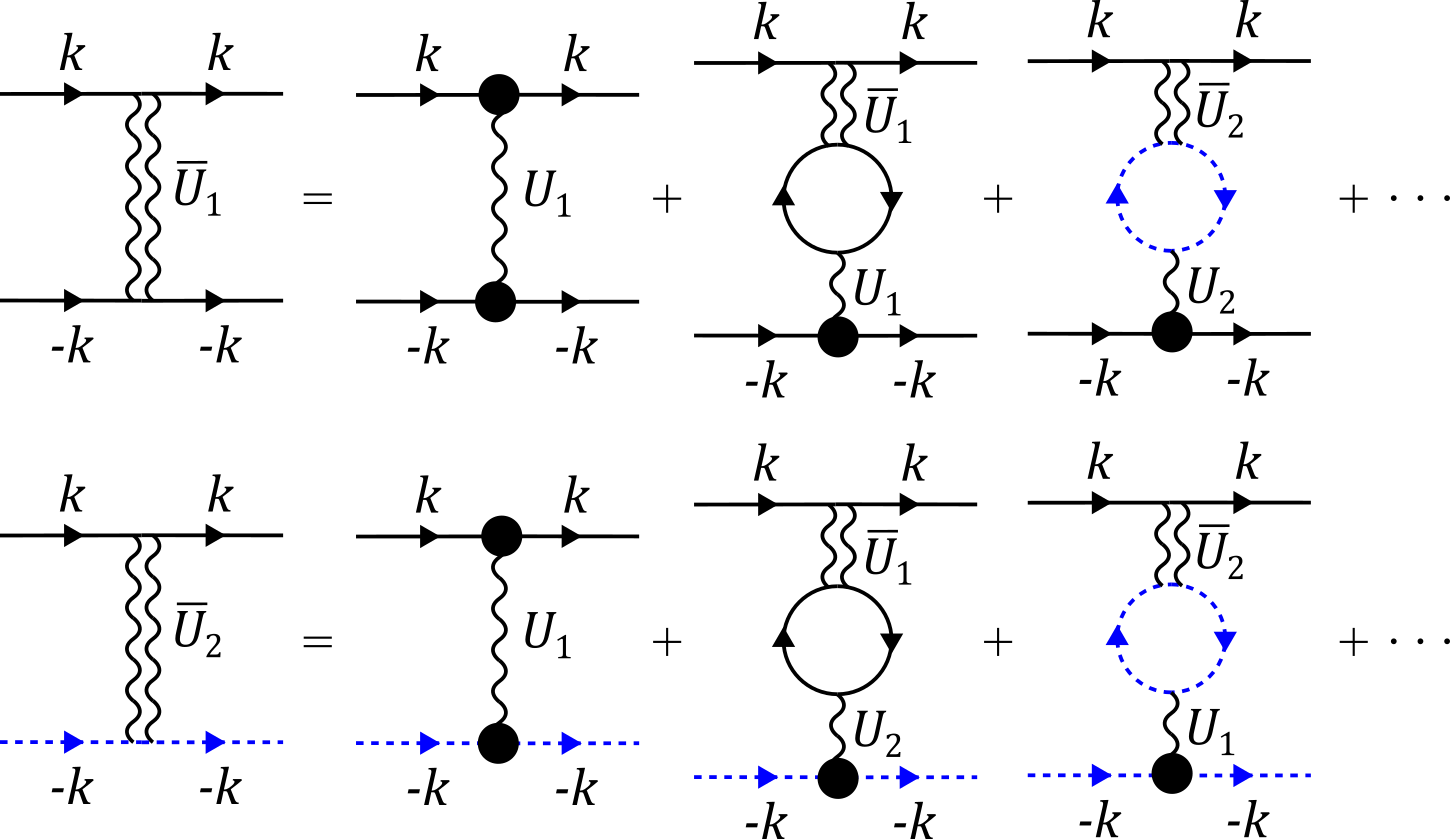}
			\centering
			\caption{Diagrammatic representation of the coupled
equations for the renormalized interactions ${\bar U}_1$  and ${\bar U}_2$ at zero momentum transfer.}
\label{fig:U1U2}
\end{figure}		
	The vertex function near the onset of
		valley polarization has been analyzed in Ref. \cite{dong_23} for the case $U_3 =0$.  In this situation, both the bare and dressed $U_b =0$, and the vertex function has only the density component $U_a$.  Relevant diagrams for $U_a$, which contain polarization
$\Pi (0)$,
are shown in Fig. \ref{fig:Ua}. They include ladder series for vertex renormalization
(Fig. \ref{fig:Ua}b), which sum up into
 $\gamma = 1/(1- U_1 \Pi (0))$,
  and series of bubbles, each of which includes ladder series of vertex renormalizations, which give one factor of $\gamma$.
The bubble diagrams can be summed up directly, in ordrer-by-order analysis. A more elegant way to sum them up
  is to re-express diagrammatic series in terms of dressed $\overline{U}_2$ (same as $U_a$) and $\overline{U}_1$ and solve the set of two coupled equations.
  The equations are shown diagrammatically in Fig. \ref{fig:U1U2}.   In analytic form we have
  	\begin{align}\label{nn_17}
			\overline{U}_1 = U_1 \gamma^2 -2 U_1\Pi(0) \overline{U}_1 \gamma -2  U_2\Pi(0) \overline{U}_2 \\
			\overline{U}_2 = U_2 \gamma^2 -2 U_1\Pi(0) \overline{U}_2 -2 U_2\Pi(0) \overline{U}_1
		\end{align}
		The solutions are
		\be \label{nn_18}
		\overline{U}_\pm = \frac{\left(U_1 \pm U_2\right)  \gamma^2}{1 + 2 \Pi (0)\left( U_1 \pm U_2\right)\gamma},
		\ee
		where $\overline{U}_\pm = \overline{U}_1\pm \overline{U}_2$.
Extracting $\overline{U}_2$, we obtain~\cite{dong_23}
 		\be \label{nn_19}
		U_a = \overline{U}_2 = \frac{\gamma^2}{2} \lb \frac{U_1 + U_2}{1+2 (U_1 + U_2) \Pi(0)\gamma}  -\frac{U_1-U_2}{1+2 (U-1-U_2)\Pi(0)\gamma}  \rb,
		\ee
Substituting  $\gamma = 1/(1-U_1 \Pi (0))$, we find that (\ref{nn_19}) simplifies to
		\be
		\overline{U}_2  = \frac{U_2}{\left(1 + (U_1 + 2 U_2)\Pi (0)
\right) \left(1 - (2U_2 -  U_1)\Pi (0)
\right)}
		\label{16}
		\ee
		We see that
$\overline{U}_2$
diverges at the onset of the valley polarization order,
 which at $U_3=0$
   is at $(2U_2 -  U_1)\Pi (0)
    =1$, Eq. (\ref{9}).
     Near the onset,
		\be
\overline{U}_2
\approx \frac{1}{4 \Pi (0)
} \frac{1}{1 - (2U_2 -  U_1)\Pi  (0)
}
		\label{17}
		\ee
				\begin{figure}
			\includegraphics[width=0.5\textwidth]{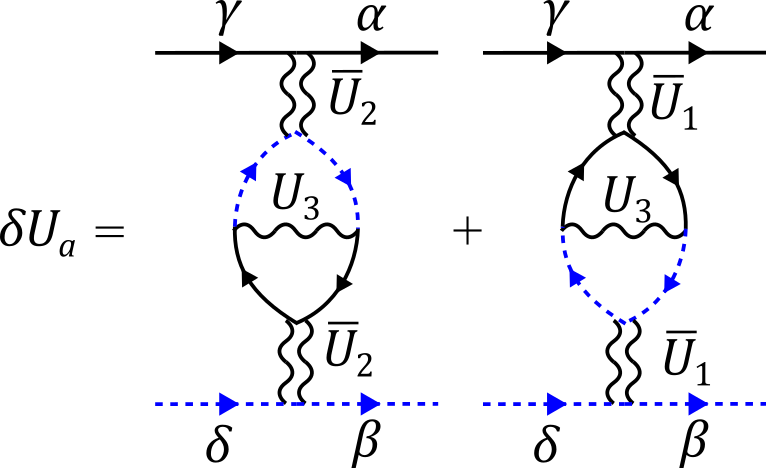}
			\centering
			\caption{Diagrams representing the correction to the effective interaction $U_a$
due to finite  $U_3$, to first order in
 $U_3$. The momenta along the upper line are approximately $(k,k)$ and the ones along the lower line are $(-k,-k)$.
}\label{fig:deltaUa}
		\end{figure}
		
		We extended the analysis of Ref.~\cite{dong_23} to include the contribution from $U_3$.
Simple experimentation with diagrammatic series to first order in $U_3$ shows that $U_b$ is finite but not singular near the onset of valley polarization. The relevant terms are
 the two extra contributions to $U_a$, which we show in
 Fig. \ref{fig:deltaUa}.  They contain $U_3 \Pi^2 (0)$ multiplied by either fully dressed
 $\overline{U}^2_2$ or
 fully dressed
 $\overline{U}^2_1$.
  With these terms,
		\be
		U_a = \overline{U}_2
- 2 U_3 \Pi^2
(0) \left(
\overline{U}^2 _2
+
\overline{U}^2_1\right)
		\label{16a}
		\ee
 The dressed $\overline{U}_1$ is extracted from (\ref{nn_18}):
		\begin{align}
\overline{U}_1 = &  \frac{U_1}{1-(U_1 \Pi (0))^2} - \frac{2 U^2_2 \Pi (0)}{1 + U_1 \Pi (0)}
		\nonumber \\
		& \times \frac{1}{\left(1 + (2U_2 +  U_1)\Pi (0)\right)
			\left(1 - (2U_2 -  U_1)\Pi (0)\right)}
		\label{16b}
		\end{align}
		Near
the valley-polarization
 instability, the relevant term in $
 \overline{U}_1$ is the divergent  second one.
   Near the onset,
  \be
\overline{U}_1
\approx   -\frac{1}{4 \Pi (0)} \frac{1}{
			\left(1 - (2U_2 -  U_1)\Pi (0)\right)}
		\label{16bb}
		\ee
  Comparing (\ref{17}) and (\ref{16bb}), we see that both $
  \overline{U}_1$ and $
  \overline{U}_2$ are proportional to the
   susceptibility of the order parameter for the valley polarization.
		Substituting these singular terms into (\ref{16a}), we find
		\begin{align}
		&
		U_a \approx \frac{1}{4 \Pi (0)} \frac{1}{1 - (2U_2 -  U_1)\Pi (0)} \left(1 - \frac{U_3 \Pi (0)}{(1 - (2U_2 -  U_1)\Pi (0))^2}\right)
		\nonumber \\
		&\approx \frac{1}{4 \Pi (0)} \frac{1}{1 - (2U_2 -  U_1 -U_3)\Pi (0)}
		\label{16c}
		\end{align}
		Comparing with (\ref{9}), we see that the inclusion of the $U_3$ shifts the singular point, where $U_a$ diverges, to exactly the onset of the valley polarization order.
		
		Substituting
$U_a$  from (\ref{16c})  into the pairing vertex,
 we obtain
		\be
		\Gamma_{\alpha \beta; \gamma \delta} (k, k'=-k;p=k,p'=-k) \approx  \frac{1}{4 \Pi (0)}
		\frac{\delta_{\alpha \beta} \delta_{\gamma \delta}}{1 - (2U_2 -  U_1-U_3)\Pi (0)}
		\label{18}
		\ee
		This form of $\Gamma$ implies that the pairing vertex is (i) enhanced near the onset of valley polarization and (ii) can be viewed as  mediated by singular charge fluctuations. This is similar to the form of the pairing interaction near the onset of density order in single-valley systems. A similarity is only partial because, in single-valley systems, boson-mediated pairing interaction is attractive. In contrast, here the pairing interaction is repulsive (positive) and, although strong, does not lead to superconductivity. One needs to add a magnetic field or spin-orbit coupling to obtain an unconventional superconductivity with frequency-dependent gap function $\Delta (\omega_m)$ that changes sign at a finite Matsubara frequency (Ref, \cite{dong_23}).
		
		\subsection{Intra-valley ferromagnetism}
		We recall that the instability towards a ferromagnetic order within a valley occurs at $1 = U_1 \Pi (0)$ if we neglect $U_3$, and at $1 = (U_1 \pm U_3) \Pi (0)$ if we keep $U_3$.
 In the last case,  the two instabilities are towards ferromagnetic and antiferromagnetic order between valleys.
		
		To obtain the pairing vertex $\Gamma$, we again need to include the renormalizations with polarization bubble $\Pi (0)$ and check whether
 there are contributions that become singular near a transition to ferromagnetism.
		Without $U_3$, this is not the case as the renormalizations are the same as in the previous section, i.e., the pairing interaction is mediated by charge fluctuations and  contains the dressed $
\overline{U}_2$. The latter  diverges at the valley polarization transition, but not at the transition toward a ferromagnetic order.

 To verify whether there is any enhanced interaction near the onset of intra-valley ferromagnetism, we then need to include the intra-valley scattering $U_3$ into consideration and analyze the structure of the terms in the pairing interaction that scale with $U_3$. Because  $U_3$ is much smaller than $U_1$ and $U_2$, we first we consider these terms to leading order in $U_3$.
 Specifically, we verify whether there exists a component of the pairing interaction that scales with $U_3$ and diverges at the onset of valley ferromagnetism at $U_1 \Pi(0) =1$.
 		\begin{figure}
			\includegraphics[width=0.9\textwidth]{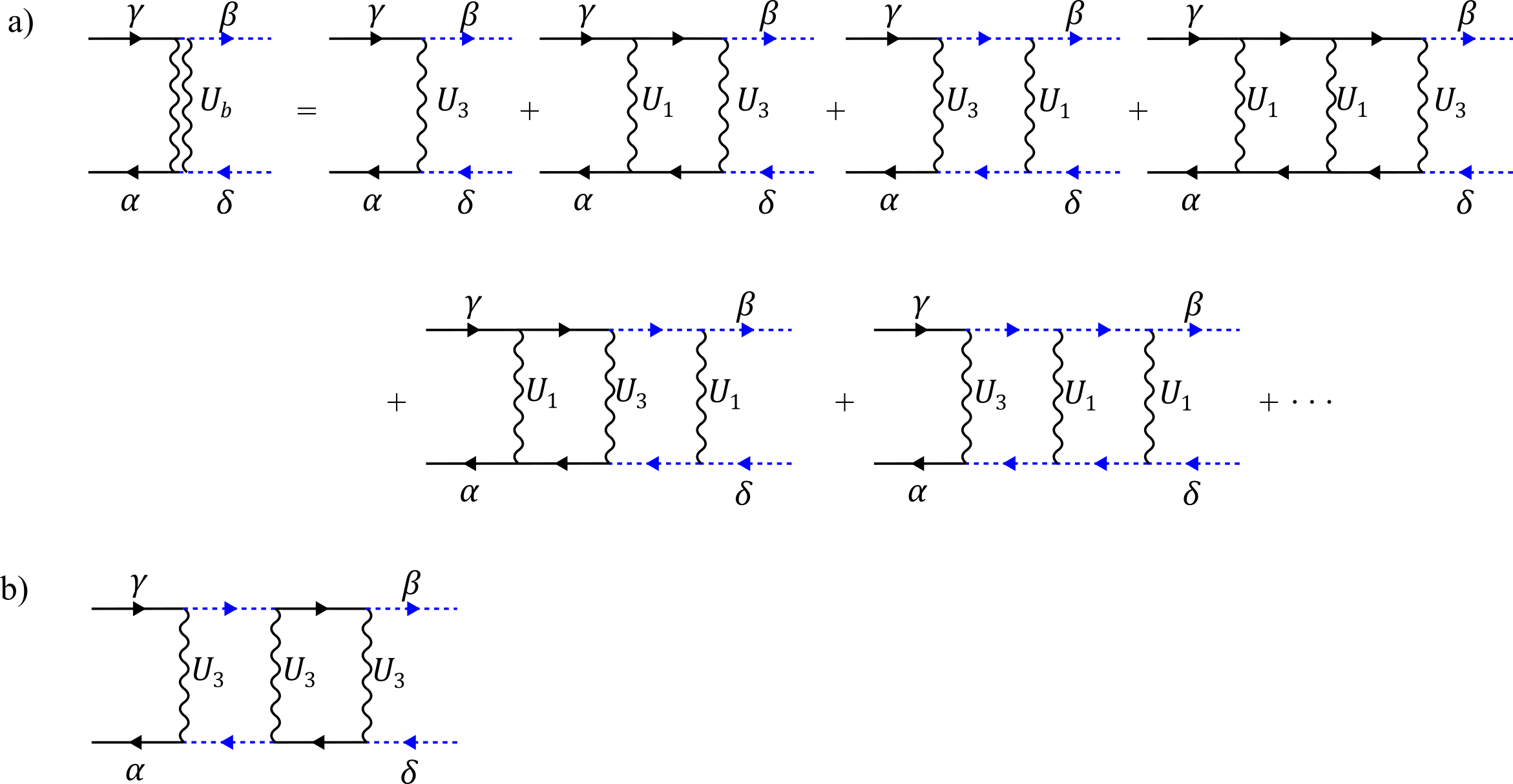}
			\centering
			\caption{Diagrammatic representation of the interaction $U_b$. Panel (a) -- - diagrams to first order in $U_3$.
 The series  yield $U_3 (1 + 2 U_1 \Pi (0) + 3 U^2_1 \Pi^2 (0) + ...)$ and sum up into $U_3/(1- U_1 \Pi (0) )^2 = \gamma^2$.  Panel (b) -- an example of a ladder diagram of higher order in
  $U_3$.}\label{fig:U3}
		\end{figure}

 We analyze both $U_a$ and $U_b$.
 The component $U_b$ equals $U_3$ at the bare level. On inspecting  perturbation series, we find
  the one, shown in Fig. \ref{fig:U3} a), that becomes singular at the onset of intera-valley ferromagnetism.
  It represents ladder insertions of intra-valley density-density interaction $U_1$ on {\it both} sides of the intra-valley scattering.
 Collecting these contributions, we find
		\be
		U_b = U_3 \gamma^2 = \frac{U_3}{(1- U_1 \Pi(0))^2}
		\label{19}
		\ee
		We see that $U_b$ diverges at the onset of ferromagnetism. Notice,  however, that the  divergent piece scales as the square of the susceptibility of the ferromagnetic order parameter. We show below that this is an artifact of keeping only the terms to leading order in $U_3$.
		
		There is also a singular contribution to $U_a$ near the onset of ferromagnetism.  The corresponding
 term is $-2 U_3 \Pi (0) \overline{U}^2_1$
 in (\ref{16a}), but now we keep only the first term  in
 the expression for $
 \overline{U}_1$ in (\ref{16b}), as it
  contains $1/(1- U_1 \Pi(0))$ and diverges when $U_1 \Pi (0) =1$.
 We then obtain
		\be
		U_a \approx - \frac{U_3}{2} \frac{(U_1 \Pi (0))^2}{(1- U_1 \Pi (0))^2} \approx - \frac{U_3}{2(1- U_1 \Pi (0))^2} 
		\label{20}
		\ee
		Substituting singular parts of $U_a$ and $U_b$ from (\ref{19}) and (\ref{20}) into the pairing vertex, we obtain
		\be
		\Gamma_{\alpha \beta; \gamma \delta} (k,-k;k,-k)
		 = - \frac{U_3}{(1- U_1 \Pi (0))^2}
		\left( \delta_{\alpha \gamma} \delta_{\beta \delta} + \frac{1}{2}
		{\vec \sigma}_{\alpha \gamma}\cdot {\vec \sigma}_{\beta \delta}\right)
		\label{21}
		\ee
		
		We see that
$\Gamma_{\alpha \beta; \gamma \delta} (k,-k;k,-k)
$ does diverge 
 at the onset of intra-valley ferromagnetism , but it is quadratic in $1/(1- U_1 \Pi(0
 ))$, and its density and spin components are comparable, i.e., the divergent $\Gamma_{\alpha \beta; \gamma \delta} (k,-k;k,-k)
 $ cannot be viewed as coming solely from spin fluctuations.

We now show
that this  result is an artifact of
restricting to  linear order in $U_3$, as
 at this level we
  do not distinguish between inter-valley ferromagnetism and inter-valley antiferromagnetism, i.e., between instabilities towards FM$^{+}$ and FM$^{-}$. From the analysis in the previous section and from general reasoning, it is natural to expect that higher-order terms in $U_3$ replace  $(1- U_1 \Pi (0))^2$ in the denominator in (\ref{21}) by $(1- U_1 \Pi (0))^2 - (U_3 \Pi(0))^2 = (1- (U_1 + U_3) \Pi (0))(1-(U_1-U_3) \Pi(0))$, such that the pairing interaction becomes singular right at the onset of either FM$^{+}$ or FM$^{-}$
   and scales with the corresponding susceptibility.  We now verify whether the divergent component comes from
   spin fluctuations.
		
\begin{figure}
			\includegraphics[width=0.6\textwidth]{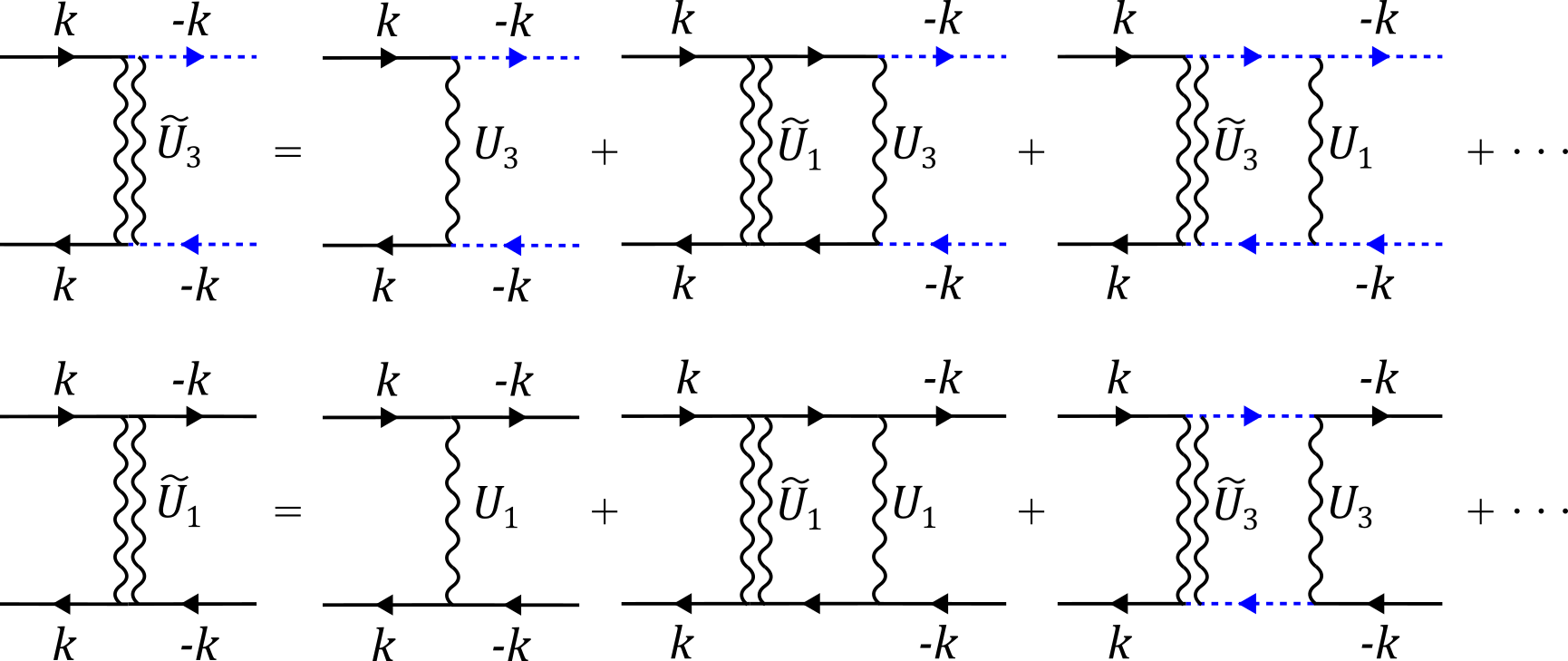}
			\centering
			\caption{Diagrammatic representation of the coupled
  equations for the dressed ${\tilde U}_3$ (the $U_b$ component of the pairing  vertex) and the dressed
   ${\tilde U}_1$. The latter is different from $\overline{U}_1$ in Fig. \protect\ref{fig:U1U2} as in these series the diagrams
     that contribute to $\overline{U}_1$ are not expressed solely in terms of $\Pi (0)$.
    }\label{fig:U1U3}
		\end{figure}

We start with $U_b$.  A simple experimentation shows that the relevant diagrams form ladder
 series shown in Fig. \ref{fig:U3}b.   There are other seemingly relevant diagrams, like the ones which we earlier  included in the renormalization of $U_1$ into $\overline{U}_1$.  However, one can verify that these diagrams are not expressed in terms of $\Pi (0)$ and by this reasons are irrelevant to our analysis.
  An infinite series of ladder diagrams can be summed up in manner similar to how it was done for $\overline{U}_1$ and $\overline{U}_2$ -- by introducing dressed vertices and re-expressing infinite series as the set of coupled equations for the dressed vertices.  We show this diagrammatically in Fig. \ref{fig:U1U3}.
In analytic form, the equations  for dressed ${\tilde U}_3 = U_b$ and the corresponding
 ${\tilde U}_1$ (different from $\overline{U}_1$ in Fig. \protect\ref{fig:U1U2} and Eq. {\ref{nn_17}) are
 		\begin{align}\label{eq:Dyson U1Ue}
			\tilde{U}_3 = U_3 +\tilde{U}_1 \Pi(0) U_3 + \tilde{U}_3 \Pi(0) U_1\\
			\tilde{U}_1 = U_1 +\tilde{U}_1 \Pi(0) U_1 + \tilde{U}_3 \Pi(0) U_3
		\end{align}
 Solving for $\tilde{U}_3 = U_b$, we obtain
		\be \label{eq:Ub_1}
		U_b = \frac{1}{2} \lb \lp U_1+U_3 \rp \gamma_+ - \lp U_1-U_3 \rp \gamma_-  \rb
		\ee
		where
\be
		\gamma_\pm = \frac{1}{ 1-\lp U_1 \pm U_3\rp \Pi(0)}
		\label{eq:Ua_for_U3neq0_a}
		\ee

The component $U_a$ is obtained in a similar way. We skip the details and present the result: $U_a$ is given by almost the same formula as (\ref{nn_19}), but now $(U_1+U_2)$ is multiplied by $\gamma_+$ and $(U_1-U_2)$ by $\gamma_-$, instead of the $\gamma$'s in Eq.\eqref{nn_19}.  In explicit form,
 		\be \label{nn_19a}
		U_a = \frac{1}{2} \lb \frac{(U_1 + U_2)\gamma^2_+}{1+2 (U_1 + U_2) \Pi(0)\gamma_+}  -\frac{(U_1-U_2) \gamma^2_-}{1+2 (U-1-U_2)\Pi(0)\gamma_-}  \rb,
		\ee
Using these results, we can analyze  superconductivity near the onset of each of the two instabilities, FM$^{+}$ and FM$^{-}$
		\begin{itemize}
			\item
			Near FM$^{+}$ where $\gamma_+\rightarrow \infty$,  we find  		\be
			U_a \approx \frac{\gamma_+}{4\Pi(0)} , \quad U_b \approx \frac{\gamma_+}{2\Pi(0)}
			\ee
			The
 pairing vertex $\Gamma_{\alpha \beta; \gamma \delta} (k,-k;k,-k)$ is
			\bea \label{eq:Gamma FM+}
			\Gamma_{\alpha \beta; \gamma \delta} (k,-k;k,-k)
			&& = - \frac{1}{4\Pi(0)} \frac{\vec \sigma_{\alpha\beta}\cdot \vec \sigma_{\gamma\delta}}{1 - (U_1 + U_3) \Pi (0)}
			\eea
			We see that the singular pairing
  vertex has only
 the  spin component,
  as one could anticipate, and scales linearly with $1/(1 - (U_1 + U_3) \Pi (0))$, i.e., it is proportional to the susceptibility of bosonic excitations associated with inter-valley ferromagnetism. The sign of the pairing interaction in the singlet and the triplet channel is determined by the sign of $\Gamma$ convoluted with $(\sigma^y_{\alpha \beta})(\sigma^y_{\gamma \delta})$ for singlet and $(\sigma^x_{\alpha \beta})(\sigma^x_{\gamma \delta})$ for triplet. One can also read off the sign of interaction using $\sigma\cdot \sigma=\pm1$ for spin-triplet/singlet. We find that the pairing interaction is repulsive (positive) for spin-singlet pairing and attractive (negative)  for spin-triplet pairing. We, therefore,   predict that near the onset of FM$^{+}$ the system becomes unstable against superconductivity in the spin-triplet valley-singlet channel.
			\item
			Near  FM$^{-}$ instability,  where $\gamma_-\rightarrow \infty$, we have
			\be \label{eq:Gamma FM-}
			\Gamma_{\alpha \beta; \gamma \delta}(k,-k;k,-k)
			= \frac{\gamma_-}{4\Pi(0)} \vec\sigma_{\alpha\beta}\cdot \vec \sigma_{\gamma\delta}
			= \frac{1}{4\Pi(0)} \frac{\vec\sigma_{\alpha\beta}\cdot \vec\sigma_{\gamma\delta}}{1 - (U_1-U_3) \Pi (0)}
			\ee
			The pairing vertex
 is again singular  {\it only} in the spin channel and is proportional to the susceptibility of bosonic excitations associated with inter-valley antiferromagnetism. However, the sign of the vertex function is different from the one near FM$^+$. As the result, the attraction is now in the spin-singlet, valley-triplet channel, while the interaction in the spin-triplet channel is repulsive. We, therefore, predict that near the onset of FM$^{-}$ the system becomes unstable against superconductivity in the
 spin-singlet, valley-triplet channel.
		\end{itemize}
		
		We also emphasize that near each instability,  the overall factor $U_3$ in the pairing interaction cancels out with $U_3$ in the denominator. Hence the dimensionless pairing coupling does not contain $U_3$. The corresponding $T_c$ still scales with $U_3$, as the latter sets the width of the range where the interaction is independent of $U_3$. Still,  there is no $1/U_3$ dependence in the exponent for $T_c$.
		
		\begin{figure}
			\includegraphics[width=0.6\textwidth]{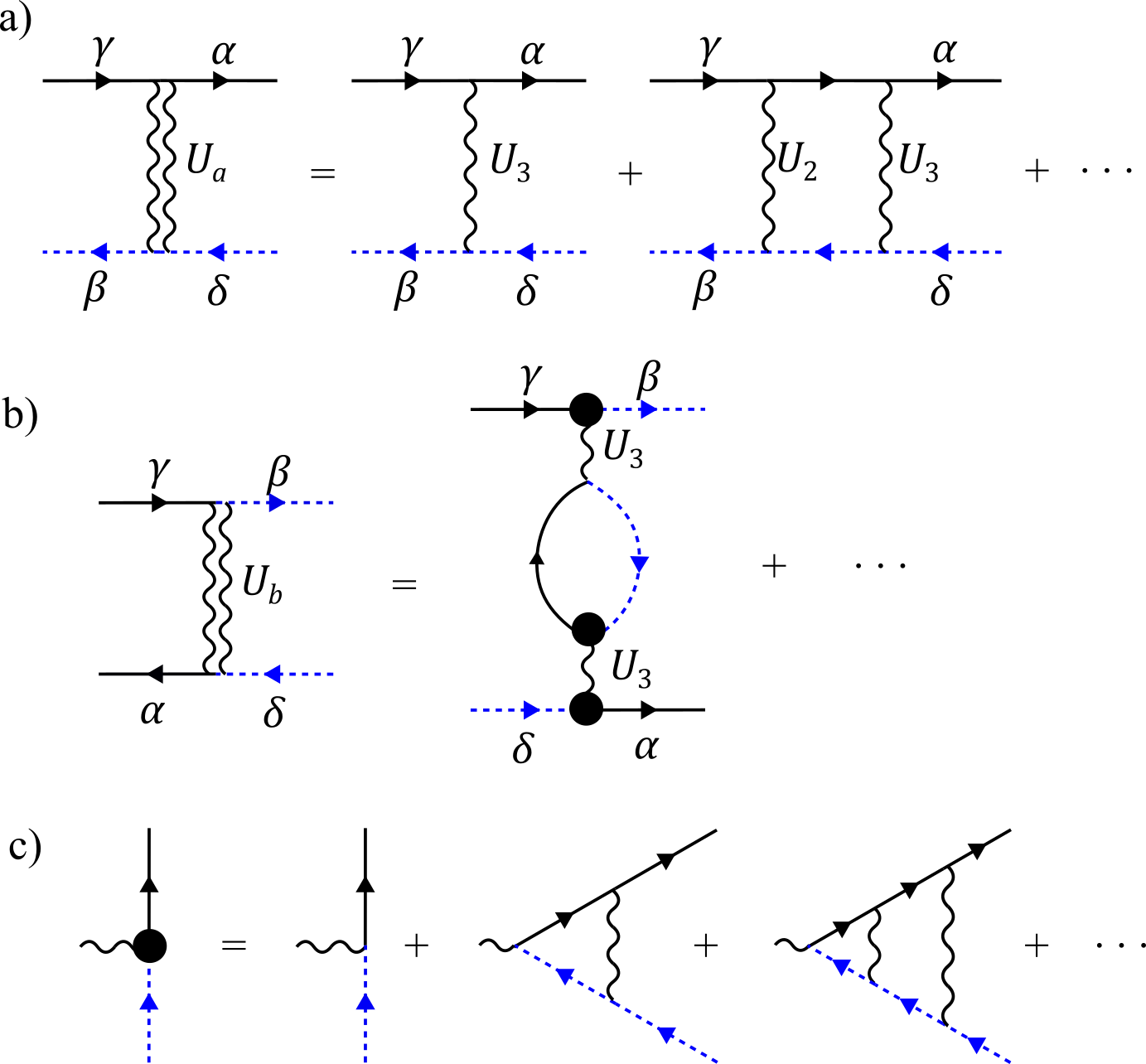}
			\centering
			\caption{Diagrammatic representation of
$U_a$ (panel (a))  and $U_b$ (panel (b)) near SDW and CDW transitions with a finite $\vec Q =\K-\KK$.  The diagrams for the dressed vertex (a black dot) are shown in panel (c).
   }\label{fig:Q}
		\end{figure}
	
	\subsection{CDW and SDW transitions at ${\vec Q} = {\vec K} - {\vec K}^{'}$}
		
		We now analyze the pairing interaction near the two transitions at a finite ${\vec Q} = \K - \KK$ -- a CDW and an SDW instability.  We again compute the two components of the pairing
vertex $\Gamma_{\alpha \beta; \gamma \delta} (k, k'=-k,p,p'=-p)$, but now we choose terms with $\Pi (Q) = \Pi (k+p)$, where $k$ and $p$ are near two Dirac points in the same valley.
 By inspecting the diagrammatic series, it becomes apparent that for $U_a$, the terms in question originate from the diagrammatic series depicted in
 Fig. \ref{fig:Q} a). This series can be interpreted as either a ladder series in $U_2$ or as a series of maximally crossed diagrams, depending on how one chooses to represent the interaction $U_2$ diagrammatically. Summing up these series, we obtain
		\be
		U_a =
\frac{U_2}{1 - U_2 \Pi (k+p)}
		\label{22}
		\ee
		
		Relevant diagrams for $U_b$ are shown in Fig. \ref{fig:Q} b.
 The dressed vertex ${\bar \gamma}$ in these series is presented in Fig. \ref{fig:Q} c). In analytic form,
 ${\bar \gamma} = 1/(1- U_2 \Pi (Q))$.  On top of this, there are insertions of bubbles, again made out of fermions from different valleys.  Summing up the bubbles and dressing each bubble and the two side vertices by $\Gamma$, we obtain
		\be
		U_b = {\bar \gamma}^2 \frac{U_3}{ 1 - 2 U_3 \Gamma \Pi (Q)} = \frac{U_3}{(1- U_2 \Pi (Q))(1- (U_2-2 U_3) \Pi (Q))}
		\label{23}
		\ee
		Substituting into the pairing vertex,
we obtain
		\be
		\Gamma_{\alpha \beta; \gamma \delta} (k,-k;k,-k)
		 =
		\frac{1}{1- U_2 \Pi (Q)} \left(U_2 \delta_{\alpha \gamma} \delta_{\beta \delta} - \frac{U_3}{1-(U_2-2U_3)\Pi (Q)} \delta_{\alpha \delta} \delta_{\beta \gamma}\right)
		\label{24}
		\ee
		This vertex can be equivalently re-expressed as
		\be
		\Gamma_{\alpha \beta; \gamma \delta} (k,-k;k,-k)
		=
		\frac{U_2}{2} \frac{{\vec \sigma}_{\alpha \delta}\cdot {\vec \sigma}_{\beta \gamma}}{1- U_2 \Pi (Q)}
		+ \frac{U_2 - 2 U_3}{2} \frac{\delta_{\alpha \delta} \delta_{\beta \gamma}}{1- (U_2-2U_3) \Pi (Q)}
		\label{26}
		\ee
		Note that the spin indices, combined into $\delta-$ and $\sigma-$functions, are for momenta
$k$ an $-p$,
like in the polarization bubble $\Pi (k+p) = \Pi ({\vec Q})$.  We see that the spin component of the vertex function is enhanced near an SDW transition at $1= U_2 \Pi ({\vec Q})$ and the charge component is enhanced near a CDW transition at $1= (U_2-2U_3) \Pi ({\vec Q})$. Near each of these two transitions, an effective pairing interaction can be viewed as mediated by fluctuations of SDW or CDW order parameters.
		
		To get the sign of the effective pairing interaction, we again convolute $\Gamma_{\alpha \beta; \gamma \delta}$ with $(\sigma^y_{\alpha \beta})(\sigma^y_{\gamma \delta})$ for spin-singlet and
		$(\sigma^x_{\alpha \beta})(\sigma^x_{\gamma \delta})$ for spin-triplet.  We find that near an SDW transition, both spin-singlet and spin-triplet components are positive (repulsive). Hence the interaction mediated by soft SDW fluctuations does not  give rise to superconductivity with a momentum-independent gap function.  Near a CDW transition, the effective interaction is repulsion in the spin-triplet channel, but attractive in the spin-singlet, valley triplet  channel.
		
		A comment is in order. In our analysis, we approximated intra-valley and inter-valley interactions as constants. As a consequence, we neglected a potential momentum dependence of the pairing interaction at momenta of order $k_F$ and treated the gap function as $\Delta_{\alpha\beta} = \Delta T^s_{\alpha\beta} T^v$, where $T^{s}$ and $T^v$ are the spin and valley Pauli matrices.  For a generic momentum-dependent pairing interaction, the gap function $\Delta_{\alpha\beta} = \Delta(k) T^s_{\alpha\beta} T^v$. For the pairing, mediated by CDW or SDW fluctuations, the pairing vertex couples $\Delta (k)$ and $\Delta (-k')$, where both $k$ and $-k'$ are near $\K$. This allows, at least in principle, two types of solutions with respect to the deviation from the center of the Fermi surface at $\K$,
which we label as ${\tilde k}$: an even parity solution $\Delta ({\tilde k}) = \Delta (-{\tilde k})$ and  an odd parity solution $\Delta ({\tilde k}) = -\Delta (-{\tilde k})$. For a constant $U_3$, only an even-parity solution is possible, but if the dressed $U_3 
(\vec p)$ varies substantially at  
$ |\vec p
 -{\vec Q}| \sim k_F$
 , an odd-parity solution is possible. For odd-parity pairing, it is natural to expect that the effective pairing vertex has an opposite sign compared to the even-parity one. Then the pairing interaction mediated by CDW fluctuations is attractive in the spin-triplet channel and repulsive in the spin-singlet channel. In contrast, the pairing interaction mediated by SDW fluctuations is attractive in both spin-triplet and spin-singlet channels.
		
		The same separation holds for the case when there are two Fermi surfaces near both $\K$ and $\KK$, one inside the other.  In this case, one can have either a sign-preserving or a sign-changing gap between the inner and outer hole pockets.
		
		This last scenario has been studied in detail for  RTG \cite{chatterjee2022inter}, and both spin-triplet and spin-singlet superconducting states have been argued to develop near the onset of SDW (``spin-polarized IVC'' states, in the nomenclature of Ref.\cite{chatterjee2022inter}).
		
		\section{Conclusions}
		This work presents a general framework to describe superconductivity triggered by correlated electronic orders in a system of interacting fermions in graphene-like bands with Fermi pockets near $\K$ and $\KK$ points in the Brillouin zone. This model, while general enough, is argued to mimic well the physics of 
		graphene multilayers
		in a displacement field. We explore four possible ordered states: valley polarization, intra-valley ferromagnetism (this order further splits into an inter-valley ferromagnetism where spin polarizations in two valleys are parallel 
		and an inter-valley antiferromagnetism where spin polarizations in two valleys are antiparallel), and CDW or SDW order at momenta ${\vec Q} = \K - \KK$, also known in the literature as intervalley coherence (IVC) orders of a spin-singlet and spin-triplet type.  Electron interactions are described by a Hamiltonian that includes a density-density interaction within a single valley ($U_1)$, an interaction between fermion densities in different valleys ($U_2$), and an inter-valley exchange interaction that involves inter-valley scattering ($U_3$).  We
		 five independent  conditions for the onset of these orders
		and considered superconductivity near each of the ordered states.  We found that the effective pairing interaction is enhanced in each case.  Near a valley polarization and CDW instability, the interaction can be viewed as mediated by soft fluctuations of the corresponding density (charge) order parameter.  The enhanced interaction is repulsive near valley polarization and attractive in the spin-singlet/valley-triplet channel near the onset of the CDW order. Near an onset of the SDW order, the enhanced pairing interaction can be viewed as mediated by soft spin fluctuations with momenta near ${\vec Q}=\vec K-\vec K'$. This interaction is, however, repulsive and as such does not give rise to superconductivity
		in s-wave channels. 
		We showed that  the pairing interaction near the onset of a ferromagnetic order within  and between valleys (FM$^{+}$ order) and one near the onset of a ferromagnetic order within a valley and antiferromagnetic order between valleys (FM$^{-}$ order) is mediated by spin fluctuations and is attractive in the 
spin-triplet/valley-singlet
 channel for FM$^{+}$ and in the
 spin-singlet/valley triplet  channel for FM$^{-}$. We argued that to demonstrate this one has to sum up infinite series in the intra-valley scattering $U_3$. In both cases, the pairing interaction scales with  $U_3$ at some distance from a magnetic transition but becomes independent on $U_3$  near the onset of the transition.  Consequently, $T_c$ at the onset of FM$^{+}$  or FM$^{-}$ order does not contain $1/U_3$ in the exponent
   (i.e., is not exponentially small in $1/U_3$).
	
These results
		substantiate the notion of graphene systems offering a versatile platform to realize and explore a wide range of possible scenarios for unconventional superconductivity driven by electron interactions.
		
		\paragraph*{\bf{Acknowledgment}} We thank E. Berg,  D. Efetov,  A. MacDonald,  and  A. Young for fruitful discussions.  The work by L.L. was supported by the Science and Technology Center for Integrated Quantum Materials, National Science Foundation Grant No. DMR1231319, and Army Research Office Grant No. W911NF-18-1-0116. The work by A.V.C. was supported by U.S. Department of Energy, Office of Science, Basic Energy Sciences, under Award No. DE-SC0014402.
\bibliography{ref_1}

\begin{thebibliography}{48}%
\makeatletter
\providecommand \@ifxundefined [1]{%
 \@ifx{#1\undefined}
}%
\providecommand \@ifnum [1]{%
 \ifnum #1\expandafter \@firstoftwo
 \else \expandafter \@secondoftwo
 \fi
}%
\providecommand \@ifx [1]{%
 \ifx #1\expandafter \@firstoftwo
 \else \expandafter \@secondoftwo
 \fi
}%
\providecommand \natexlab [1]{#1}%
\providecommand \enquote  [1]{``#1''}%
\providecommand \bibnamefont  [1]{#1}%
\providecommand \bibfnamefont [1]{#1}%
\providecommand \citenamefont [1]{#1}%
\providecommand \href@noop [0]{\@secondoftwo}%
\providecommand \href [0]{\begingroup \@sanitize@url \@href}%
\providecommand \@href[1]{\@@startlink{#1}\@@href}%
\providecommand \@@href[1]{\endgroup#1\@@endlink}%
\providecommand \@sanitize@url [0]{\catcode `\\12\catcode `\$12\catcode
  `\&12\catcode `\#12\catcode `\^12\catcode `\_12\catcode `\%12\relax}%
\providecommand \@@startlink[1]{}%
\providecommand \@@endlink[0]{}%
\providecommand \url  [0]{\begingroup\@sanitize@url \@url }%
\providecommand \@url [1]{\endgroup\@href {#1}{\urlprefix }}%
\providecommand \urlprefix  [0]{URL }%
\providecommand \Eprint [0]{\href }%
\providecommand \doibase [0]{https://doi.org/}%
\providecommand \selectlanguage [0]{\@gobble}%
\providecommand \bibinfo  [0]{\@secondoftwo}%
\providecommand \bibfield  [0]{\@secondoftwo}%
\providecommand \translation [1]{[#1]}%
\providecommand \BibitemOpen [0]{}%
\providecommand \bibitemStop [0]{}%
\providecommand \bibitemNoStop [0]{.\EOS\space}%
\providecommand \EOS [0]{\spacefactor3000\relax}%
\providecommand \BibitemShut  [1]{\csname bibitem#1\endcsname}%
\let\auto@bib@innerbib\@empty
\bibitem [{\citenamefont {de~la Barrera}\ \emph {et~al.}(2021)\citenamefont
  {de~la Barrera}, \citenamefont {Aronson}, \citenamefont {Zheng},
  \citenamefont {Watanabe}, \citenamefont {Taniguchi}, \citenamefont {Ma},
  \citenamefont {Jarillo-Herrero},\ and\ \citenamefont
  {Ashoori}}]{de2021cascade}%
  \BibitemOpen
  \bibfield  {author} {\bibinfo {author} {\bibfnamefont {S.~C.}\ \bibnamefont
  {de~la Barrera}}, \bibinfo {author} {\bibfnamefont {S.}~\bibnamefont
  {Aronson}}, \bibinfo {author} {\bibfnamefont {Z.}~\bibnamefont {Zheng}},
  \bibinfo {author} {\bibfnamefont {K.}~\bibnamefont {Watanabe}}, \bibinfo
  {author} {\bibfnamefont {T.}~\bibnamefont {Taniguchi}}, \bibinfo {author}
  {\bibfnamefont {Q.}~\bibnamefont {Ma}}, \bibinfo {author} {\bibfnamefont
  {P.}~\bibnamefont {Jarillo-Herrero}},\ and\ \bibinfo {author} {\bibfnamefont
  {R.}~\bibnamefont {Ashoori}},\ }\href@noop {} {\bibfield  {journal} {\bibinfo
   {journal} {arXiv preprint arXiv:2110.13907}\ } (\bibinfo {year}
  {2021})}\BibitemShut {NoStop}%
\bibitem [{\citenamefont {Seiler}\ \emph {et~al.}(2022)\citenamefont {Seiler},
  \citenamefont {Geisenhof}, \citenamefont {Winterer}, \citenamefont
  {Watanabe}, \citenamefont {Taniguchi}, \citenamefont {Xu}, \citenamefont
  {Zhang},\ and\ \citenamefont {Weitz}}]{seiler2022quantum}%
  \BibitemOpen
  \bibfield  {author} {\bibinfo {author} {\bibfnamefont {A.~M.}\ \bibnamefont
  {Seiler}}, \bibinfo {author} {\bibfnamefont {F.~R.}\ \bibnamefont
  {Geisenhof}}, \bibinfo {author} {\bibfnamefont {F.}~\bibnamefont {Winterer}},
  \bibinfo {author} {\bibfnamefont {K.}~\bibnamefont {Watanabe}}, \bibinfo
  {author} {\bibfnamefont {T.}~\bibnamefont {Taniguchi}}, \bibinfo {author}
  {\bibfnamefont {T.}~\bibnamefont {Xu}}, \bibinfo {author} {\bibfnamefont
  {F.}~\bibnamefont {Zhang}},\ and\ \bibinfo {author} {\bibfnamefont {R.~T.}\
  \bibnamefont {Weitz}},\ }\href@noop {} {\bibfield  {journal} {\bibinfo
  {journal} {Nature}\ }\textbf {\bibinfo {volume} {608}},\ \bibinfo {pages}
  {298} (\bibinfo {year} {2022})}\BibitemShut {NoStop}%
\bibitem [{\citenamefont {Zhou}\ \emph {et~al.}(2022)\citenamefont {Zhou},
  \citenamefont {Holleis}, \citenamefont {Saito}, \citenamefont {Cohen},
  \citenamefont {Huynh}, \citenamefont {Patterson}, \citenamefont {Yang},
  \citenamefont {Taniguchi}, \citenamefont {Watanabe},\ and\ \citenamefont
  {Young}}]{zhou2022isospin}%
  \BibitemOpen
  \bibfield  {author} {\bibinfo {author} {\bibfnamefont {H.}~\bibnamefont
  {Zhou}}, \bibinfo {author} {\bibfnamefont {L.}~\bibnamefont {Holleis}},
  \bibinfo {author} {\bibfnamefont {Y.}~\bibnamefont {Saito}}, \bibinfo
  {author} {\bibfnamefont {L.}~\bibnamefont {Cohen}}, \bibinfo {author}
  {\bibfnamefont {W.}~\bibnamefont {Huynh}}, \bibinfo {author} {\bibfnamefont
  {C.~L.}\ \bibnamefont {Patterson}}, \bibinfo {author} {\bibfnamefont
  {F.}~\bibnamefont {Yang}}, \bibinfo {author} {\bibfnamefont {T.}~\bibnamefont
  {Taniguchi}}, \bibinfo {author} {\bibfnamefont {K.}~\bibnamefont
  {Watanabe}},\ and\ \bibinfo {author} {\bibfnamefont {A.~F.}\ \bibnamefont
  {Young}},\ }\href@noop {} {\bibfield  {journal} {\bibinfo  {journal}
  {Science}\ }\textbf {\bibinfo {volume} {375}},\ \bibinfo {pages} {774}
  (\bibinfo {year} {2022})}\BibitemShut {NoStop}%
\bibitem [{\citenamefont {Zhang}\ \emph {et~al.}(2022)\citenamefont {Zhang},
  \citenamefont {Polski}, \citenamefont {Thomson}, \citenamefont
  {Lantagne-Hurtubise}, \citenamefont {Lewandowski}, \citenamefont {Zhou},
  \citenamefont {Watanabe}, \citenamefont {Taniguchi}, \citenamefont {Alicea},\
  and\ \citenamefont {Nadj-Perge}}]{zhang2022spin}%
  \BibitemOpen
  \bibfield  {author} {\bibinfo {author} {\bibfnamefont {Y.}~\bibnamefont
  {Zhang}}, \bibinfo {author} {\bibfnamefont {R.}~\bibnamefont {Polski}},
  \bibinfo {author} {\bibfnamefont {A.}~\bibnamefont {Thomson}}, \bibinfo
  {author} {\bibfnamefont {{\'E}.}~\bibnamefont {Lantagne-Hurtubise}}, \bibinfo
  {author} {\bibfnamefont {C.}~\bibnamefont {Lewandowski}}, \bibinfo {author}
  {\bibfnamefont {H.}~\bibnamefont {Zhou}}, \bibinfo {author} {\bibfnamefont
  {K.}~\bibnamefont {Watanabe}}, \bibinfo {author} {\bibfnamefont
  {T.}~\bibnamefont {Taniguchi}}, \bibinfo {author} {\bibfnamefont
  {J.}~\bibnamefont {Alicea}},\ and\ \bibinfo {author} {\bibfnamefont
  {S.}~\bibnamefont {Nadj-Perge}},\ }\href@noop {} {\bibfield  {journal}
  {\bibinfo  {journal} {arXiv preprint arXiv:2205.05087}\ } (\bibinfo {year}
  {2022})}\BibitemShut {NoStop}%
\bibitem [{\citenamefont {Holleis}\ \emph {et~al.}(2023)\citenamefont
  {Holleis}, \citenamefont {Patterson}, \citenamefont {Zhang}, \citenamefont
  {Yoo}, \citenamefont {Zhou}, \citenamefont {Taniguchi}, \citenamefont
  {Watanabe}, \citenamefont {Nadj-Perge},\ and\ \citenamefont
  {Young}}]{holleis2023ising}%
  \BibitemOpen
  \bibfield  {author} {\bibinfo {author} {\bibfnamefont {L.}~\bibnamefont
  {Holleis}}, \bibinfo {author} {\bibfnamefont {C.~L.}\ \bibnamefont
  {Patterson}}, \bibinfo {author} {\bibfnamefont {Y.}~\bibnamefont {Zhang}},
  \bibinfo {author} {\bibfnamefont {H.~M.}\ \bibnamefont {Yoo}}, \bibinfo
  {author} {\bibfnamefont {H.}~\bibnamefont {Zhou}}, \bibinfo {author}
  {\bibfnamefont {T.}~\bibnamefont {Taniguchi}}, \bibinfo {author}
  {\bibfnamefont {K.}~\bibnamefont {Watanabe}}, \bibinfo {author}
  {\bibfnamefont {S.}~\bibnamefont {Nadj-Perge}},\ and\ \bibinfo {author}
  {\bibfnamefont {A.~F.}\ \bibnamefont {Young}},\ }\href@noop {} {\bibfield
  {journal} {\bibinfo  {journal} {arXiv preprint arXiv:2303.00742}\ } (\bibinfo
  {year} {2023})}\BibitemShut {NoStop}%
\bibitem [{\citenamefont {Cao}\ \emph {et~al.}(2018{\natexlab{a}})\citenamefont
  {Cao}, \citenamefont {Fatemi}, \citenamefont {Fang}, \citenamefont
  {Watanabe}, \citenamefont {Taniguchi}, \citenamefont {Kaxiras},\ and\
  \citenamefont {Jarillo-Herrero}}]{cao2018unconventional}%
  \BibitemOpen
  \bibfield  {author} {\bibinfo {author} {\bibfnamefont {Y.}~\bibnamefont
  {Cao}}, \bibinfo {author} {\bibfnamefont {V.}~\bibnamefont {Fatemi}},
  \bibinfo {author} {\bibfnamefont {S.}~\bibnamefont {Fang}}, \bibinfo {author}
  {\bibfnamefont {K.}~\bibnamefont {Watanabe}}, \bibinfo {author}
  {\bibfnamefont {T.}~\bibnamefont {Taniguchi}}, \bibinfo {author}
  {\bibfnamefont {E.}~\bibnamefont {Kaxiras}},\ and\ \bibinfo {author}
  {\bibfnamefont {P.}~\bibnamefont {Jarillo-Herrero}},\ }\href
  {https://doi.org/10.1038/nature26160} {\bibfield  {journal} {\bibinfo
  {journal} {Nature}\ }\textbf {\bibinfo {volume} {556}},\ \bibinfo {pages}
  {43–50} (\bibinfo {year} {2018}{\natexlab{a}})}\BibitemShut {NoStop}%
\bibitem [{\citenamefont {Yankowitz}\ \emph {et~al.}(2019)\citenamefont
  {Yankowitz}, \citenamefont {Chen}, \citenamefont {Polshyn}, \citenamefont
  {Zhang}, \citenamefont {Watanabe}, \citenamefont {Taniguchi}, \citenamefont
  {Graf}, \citenamefont {Young},\ and\ \citenamefont
  {Dean}}]{yankowitz2019tuning}%
  \BibitemOpen
  \bibfield  {author} {\bibinfo {author} {\bibfnamefont {M.}~\bibnamefont
  {Yankowitz}}, \bibinfo {author} {\bibfnamefont {S.}~\bibnamefont {Chen}},
  \bibinfo {author} {\bibfnamefont {H.}~\bibnamefont {Polshyn}}, \bibinfo
  {author} {\bibfnamefont {Y.}~\bibnamefont {Zhang}}, \bibinfo {author}
  {\bibfnamefont {K.}~\bibnamefont {Watanabe}}, \bibinfo {author}
  {\bibfnamefont {T.}~\bibnamefont {Taniguchi}}, \bibinfo {author}
  {\bibfnamefont {D.}~\bibnamefont {Graf}}, \bibinfo {author} {\bibfnamefont
  {A.~F.}\ \bibnamefont {Young}},\ and\ \bibinfo {author} {\bibfnamefont
  {C.~R.}\ \bibnamefont {Dean}},\ }\href@noop {} {\bibfield  {journal}
  {\bibinfo  {journal} {Science}\ }\textbf {\bibinfo {volume} {363}},\ \bibinfo
  {pages} {1059} (\bibinfo {year} {2019})}\BibitemShut {NoStop}%
\bibitem [{\citenamefont {Lu}\ \emph {et~al.}(2019)\citenamefont {Lu},
  \citenamefont {Stepanov}, \citenamefont {Yang}, \citenamefont {Xie},
  \citenamefont {Aamir}, \citenamefont {Das}, \citenamefont {Urgell},
  \citenamefont {Watanabe}, \citenamefont {Taniguchi}, \citenamefont {Zhang}
  \emph {et~al.}}]{lu2019superconductors}%
  \BibitemOpen
  \bibfield  {author} {\bibinfo {author} {\bibfnamefont {X.}~\bibnamefont
  {Lu}}, \bibinfo {author} {\bibfnamefont {P.}~\bibnamefont {Stepanov}},
  \bibinfo {author} {\bibfnamefont {W.}~\bibnamefont {Yang}}, \bibinfo {author}
  {\bibfnamefont {M.}~\bibnamefont {Xie}}, \bibinfo {author} {\bibfnamefont
  {M.~A.}\ \bibnamefont {Aamir}}, \bibinfo {author} {\bibfnamefont
  {I.}~\bibnamefont {Das}}, \bibinfo {author} {\bibfnamefont {C.}~\bibnamefont
  {Urgell}}, \bibinfo {author} {\bibfnamefont {K.}~\bibnamefont {Watanabe}},
  \bibinfo {author} {\bibfnamefont {T.}~\bibnamefont {Taniguchi}}, \bibinfo
  {author} {\bibfnamefont {G.}~\bibnamefont {Zhang}}, \emph {et~al.},\
  }\href@noop {} {\bibfield  {journal} {\bibinfo  {journal} {Nature}\ }\textbf
  {\bibinfo {volume} {574}},\ \bibinfo {pages} {653} (\bibinfo {year}
  {2019})}\BibitemShut {NoStop}%
\bibitem [{\citenamefont {Cao}\ \emph {et~al.}(2021)\citenamefont {Cao},
  \citenamefont {Rodan-Legrain}, \citenamefont {Park}, \citenamefont {Yuan},
  \citenamefont {Watanabe}, \citenamefont {Taniguchi}, \citenamefont
  {Fernandes}, \citenamefont {Fu},\ and\ \citenamefont
  {Jarillo-Herrero}}]{cao2021nematicity}%
  \BibitemOpen
  \bibfield  {author} {\bibinfo {author} {\bibfnamefont {Y.}~\bibnamefont
  {Cao}}, \bibinfo {author} {\bibfnamefont {D.}~\bibnamefont {Rodan-Legrain}},
  \bibinfo {author} {\bibfnamefont {J.~M.}\ \bibnamefont {Park}}, \bibinfo
  {author} {\bibfnamefont {N.~F.}\ \bibnamefont {Yuan}}, \bibinfo {author}
  {\bibfnamefont {K.}~\bibnamefont {Watanabe}}, \bibinfo {author}
  {\bibfnamefont {T.}~\bibnamefont {Taniguchi}}, \bibinfo {author}
  {\bibfnamefont {R.~M.}\ \bibnamefont {Fernandes}}, \bibinfo {author}
  {\bibfnamefont {L.}~\bibnamefont {Fu}},\ and\ \bibinfo {author}
  {\bibfnamefont {P.}~\bibnamefont {Jarillo-Herrero}},\ }\href@noop {}
  {\bibfield  {journal} {\bibinfo  {journal} {science}\ }\textbf {\bibinfo
  {volume} {372}},\ \bibinfo {pages} {264} (\bibinfo {year}
  {2021})}\BibitemShut {NoStop}%
\bibitem [{\citenamefont {Cao}\ \emph {et~al.}(2018{\natexlab{b}})\citenamefont
  {Cao}, \citenamefont {Fatemi}, \citenamefont {Demir}, \citenamefont {Fang},
  \citenamefont {Tomarken}, \citenamefont {Luo}, \citenamefont
  {Sanchez-Yamagishi}, \citenamefont {Watanabe}, \citenamefont {Taniguchi},
  \citenamefont {Kaxiras},\ and\ \citenamefont {et~al.}}]{cao2018correlated}%
  \BibitemOpen
  \bibfield  {author} {\bibinfo {author} {\bibfnamefont {Y.}~\bibnamefont
  {Cao}}, \bibinfo {author} {\bibfnamefont {V.}~\bibnamefont {Fatemi}},
  \bibinfo {author} {\bibfnamefont {A.}~\bibnamefont {Demir}}, \bibinfo
  {author} {\bibfnamefont {S.}~\bibnamefont {Fang}}, \bibinfo {author}
  {\bibfnamefont {S.~L.}\ \bibnamefont {Tomarken}}, \bibinfo {author}
  {\bibfnamefont {J.~Y.}\ \bibnamefont {Luo}}, \bibinfo {author} {\bibfnamefont
  {J.~D.}\ \bibnamefont {Sanchez-Yamagishi}}, \bibinfo {author} {\bibfnamefont
  {K.}~\bibnamefont {Watanabe}}, \bibinfo {author} {\bibfnamefont
  {T.}~\bibnamefont {Taniguchi}}, \bibinfo {author} {\bibfnamefont
  {E.}~\bibnamefont {Kaxiras}},\ and\ \bibinfo {author} {\bibnamefont
  {et~al.}},\ }\href {https://doi.org/10.1038/nature26154} {\bibfield
  {journal} {\bibinfo  {journal} {Nature}\ }\textbf {\bibinfo {volume} {556}},\
  \bibinfo {pages} {80–84} (\bibinfo {year}
  {2018}{\natexlab{b}})}\BibitemShut {NoStop}%
\bibitem [{\citenamefont {Jiang}\ \emph {et~al.}(2019)\citenamefont {Jiang},
  \citenamefont {Lai}, \citenamefont {Watanabe}, \citenamefont {Taniguchi},
  \citenamefont {Haule}, \citenamefont {Mao},\ and\ \citenamefont
  {Andrei}}]{jiang2019charge}%
  \BibitemOpen
  \bibfield  {author} {\bibinfo {author} {\bibfnamefont {Y.}~\bibnamefont
  {Jiang}}, \bibinfo {author} {\bibfnamefont {X.}~\bibnamefont {Lai}}, \bibinfo
  {author} {\bibfnamefont {K.}~\bibnamefont {Watanabe}}, \bibinfo {author}
  {\bibfnamefont {T.}~\bibnamefont {Taniguchi}}, \bibinfo {author}
  {\bibfnamefont {K.}~\bibnamefont {Haule}}, \bibinfo {author} {\bibfnamefont
  {J.}~\bibnamefont {Mao}},\ and\ \bibinfo {author} {\bibfnamefont {E.~Y.}\
  \bibnamefont {Andrei}},\ }\href@noop {} {\bibfield  {journal} {\bibinfo
  {journal} {Nature}\ }\textbf {\bibinfo {volume} {573}},\ \bibinfo {pages}
  {91} (\bibinfo {year} {2019})}\BibitemShut {NoStop}%
\bibitem [{\citenamefont {Saito}\ \emph
  {et~al.}(2021{\natexlab{a}})\citenamefont {Saito}, \citenamefont {Yang},
  \citenamefont {Ge}, \citenamefont {Liu}, \citenamefont {Taniguchi},
  \citenamefont {Watanabe}, \citenamefont {Li}, \citenamefont {Berg},\ and\
  \citenamefont {Young}}]{saito2021isospin}%
  \BibitemOpen
  \bibfield  {author} {\bibinfo {author} {\bibfnamefont {Y.}~\bibnamefont
  {Saito}}, \bibinfo {author} {\bibfnamefont {F.}~\bibnamefont {Yang}},
  \bibinfo {author} {\bibfnamefont {J.}~\bibnamefont {Ge}}, \bibinfo {author}
  {\bibfnamefont {X.}~\bibnamefont {Liu}}, \bibinfo {author} {\bibfnamefont
  {T.}~\bibnamefont {Taniguchi}}, \bibinfo {author} {\bibfnamefont
  {K.}~\bibnamefont {Watanabe}}, \bibinfo {author} {\bibfnamefont
  {J.}~\bibnamefont {Li}}, \bibinfo {author} {\bibfnamefont {E.}~\bibnamefont
  {Berg}},\ and\ \bibinfo {author} {\bibfnamefont {A.~F.}\ \bibnamefont
  {Young}},\ }\href@noop {} {\bibfield  {journal} {\bibinfo  {journal}
  {Nature}\ }\textbf {\bibinfo {volume} {592}},\ \bibinfo {pages} {220}
  (\bibinfo {year} {2021}{\natexlab{a}})}\BibitemShut {NoStop}%
\bibitem [{\citenamefont {Zondiner}\ \emph {et~al.}(2020)\citenamefont
  {Zondiner}, \citenamefont {Rozen}, \citenamefont {Rodan-Legrain},
  \citenamefont {Cao}, \citenamefont {Queiroz}, \citenamefont {Taniguchi},
  \citenamefont {Watanabe}, \citenamefont {Oreg}, \citenamefont {von Oppen},
  \citenamefont {Stern} \emph {et~al.}}]{zondiner2020cascade}%
  \BibitemOpen
  \bibfield  {author} {\bibinfo {author} {\bibfnamefont {U.}~\bibnamefont
  {Zondiner}}, \bibinfo {author} {\bibfnamefont {A.}~\bibnamefont {Rozen}},
  \bibinfo {author} {\bibfnamefont {D.}~\bibnamefont {Rodan-Legrain}}, \bibinfo
  {author} {\bibfnamefont {Y.}~\bibnamefont {Cao}}, \bibinfo {author}
  {\bibfnamefont {R.}~\bibnamefont {Queiroz}}, \bibinfo {author} {\bibfnamefont
  {T.}~\bibnamefont {Taniguchi}}, \bibinfo {author} {\bibfnamefont
  {K.}~\bibnamefont {Watanabe}}, \bibinfo {author} {\bibfnamefont
  {Y.}~\bibnamefont {Oreg}}, \bibinfo {author} {\bibfnamefont {F.}~\bibnamefont
  {von Oppen}}, \bibinfo {author} {\bibfnamefont {A.}~\bibnamefont {Stern}},
  \emph {et~al.},\ }\href@noop {} {\bibfield  {journal} {\bibinfo  {journal}
  {Nature}\ }\textbf {\bibinfo {volume} {582}},\ \bibinfo {pages} {203}
  (\bibinfo {year} {2020})}\BibitemShut {NoStop}%
\bibitem [{\citenamefont {Choi}\ \emph {et~al.}(2021)\citenamefont {Choi},
  \citenamefont {Kim}, \citenamefont {Peng}, \citenamefont {Thomson},
  \citenamefont {Lewandowski}, \citenamefont {Polski}, \citenamefont {Zhang},
  \citenamefont {Arora}, \citenamefont {Watanabe}, \citenamefont {Taniguchi}
  \emph {et~al.}}]{choi2021correlation}%
  \BibitemOpen
  \bibfield  {author} {\bibinfo {author} {\bibfnamefont {Y.}~\bibnamefont
  {Choi}}, \bibinfo {author} {\bibfnamefont {H.}~\bibnamefont {Kim}}, \bibinfo
  {author} {\bibfnamefont {Y.}~\bibnamefont {Peng}}, \bibinfo {author}
  {\bibfnamefont {A.}~\bibnamefont {Thomson}}, \bibinfo {author} {\bibfnamefont
  {C.}~\bibnamefont {Lewandowski}}, \bibinfo {author} {\bibfnamefont
  {R.}~\bibnamefont {Polski}}, \bibinfo {author} {\bibfnamefont
  {Y.}~\bibnamefont {Zhang}}, \bibinfo {author} {\bibfnamefont {H.~S.}\
  \bibnamefont {Arora}}, \bibinfo {author} {\bibfnamefont {K.}~\bibnamefont
  {Watanabe}}, \bibinfo {author} {\bibfnamefont {T.}~\bibnamefont {Taniguchi}},
  \emph {et~al.},\ }\href@noop {} {\bibfield  {journal} {\bibinfo  {journal}
  {Nature}\ }\textbf {\bibinfo {volume} {589}},\ \bibinfo {pages} {536}
  (\bibinfo {year} {2021})}\BibitemShut {NoStop}%
\bibitem [{\citenamefont {Rozen}\ \emph {et~al.}(2021)\citenamefont {Rozen},
  \citenamefont {Park}, \citenamefont {Zondiner}, \citenamefont {Cao},
  \citenamefont {Rodan-Legrain}, \citenamefont {Taniguchi}, \citenamefont
  {Watanabe}, \citenamefont {Oreg}, \citenamefont {Stern}, \citenamefont {Berg}
  \emph {et~al.}}]{rozen2021entropic}%
  \BibitemOpen
  \bibfield  {author} {\bibinfo {author} {\bibfnamefont {A.}~\bibnamefont
  {Rozen}}, \bibinfo {author} {\bibfnamefont {J.~M.}\ \bibnamefont {Park}},
  \bibinfo {author} {\bibfnamefont {U.}~\bibnamefont {Zondiner}}, \bibinfo
  {author} {\bibfnamefont {Y.}~\bibnamefont {Cao}}, \bibinfo {author}
  {\bibfnamefont {D.}~\bibnamefont {Rodan-Legrain}}, \bibinfo {author}
  {\bibfnamefont {T.}~\bibnamefont {Taniguchi}}, \bibinfo {author}
  {\bibfnamefont {K.}~\bibnamefont {Watanabe}}, \bibinfo {author}
  {\bibfnamefont {Y.}~\bibnamefont {Oreg}}, \bibinfo {author} {\bibfnamefont
  {A.}~\bibnamefont {Stern}}, \bibinfo {author} {\bibfnamefont
  {E.}~\bibnamefont {Berg}}, \emph {et~al.},\ }\href@noop {} {\bibfield
  {journal} {\bibinfo  {journal} {Nature}\ }\textbf {\bibinfo {volume} {592}},\
  \bibinfo {pages} {214} (\bibinfo {year} {2021})}\BibitemShut {NoStop}%
\bibitem [{\citenamefont {Pierce}\ \emph {et~al.}(2021)\citenamefont {Pierce},
  \citenamefont {Xie}, \citenamefont {Park}, \citenamefont {Khalaf},
  \citenamefont {Lee}, \citenamefont {Cao}, \citenamefont {Parker},
  \citenamefont {Forrester}, \citenamefont {Chen}, \citenamefont {Watanabe}
  \emph {et~al.}}]{pierce2021unconventional}%
  \BibitemOpen
  \bibfield  {author} {\bibinfo {author} {\bibfnamefont {A.~T.}\ \bibnamefont
  {Pierce}}, \bibinfo {author} {\bibfnamefont {Y.}~\bibnamefont {Xie}},
  \bibinfo {author} {\bibfnamefont {J.~M.}\ \bibnamefont {Park}}, \bibinfo
  {author} {\bibfnamefont {E.}~\bibnamefont {Khalaf}}, \bibinfo {author}
  {\bibfnamefont {S.~H.}\ \bibnamefont {Lee}}, \bibinfo {author} {\bibfnamefont
  {Y.}~\bibnamefont {Cao}}, \bibinfo {author} {\bibfnamefont {D.~E.}\
  \bibnamefont {Parker}}, \bibinfo {author} {\bibfnamefont {P.~R.}\
  \bibnamefont {Forrester}}, \bibinfo {author} {\bibfnamefont {S.}~\bibnamefont
  {Chen}}, \bibinfo {author} {\bibfnamefont {K.}~\bibnamefont {Watanabe}},
  \emph {et~al.},\ }\href@noop {} {\bibfield  {journal} {\bibinfo  {journal}
  {Nature Physics}\ }\textbf {\bibinfo {volume} {17}},\ \bibinfo {pages} {1210}
  (\bibinfo {year} {2021})}\BibitemShut {NoStop}%
\bibitem [{\citenamefont {Saito}\ \emph
  {et~al.}(2021{\natexlab{b}})\citenamefont {Saito}, \citenamefont {Ge},
  \citenamefont {Rademaker}, \citenamefont {Watanabe}, \citenamefont
  {Taniguchi}, \citenamefont {Abanin},\ and\ \citenamefont
  {Young}}]{saito2021hofstadter}%
  \BibitemOpen
  \bibfield  {author} {\bibinfo {author} {\bibfnamefont {Y.}~\bibnamefont
  {Saito}}, \bibinfo {author} {\bibfnamefont {J.}~\bibnamefont {Ge}}, \bibinfo
  {author} {\bibfnamefont {L.}~\bibnamefont {Rademaker}}, \bibinfo {author}
  {\bibfnamefont {K.}~\bibnamefont {Watanabe}}, \bibinfo {author}
  {\bibfnamefont {T.}~\bibnamefont {Taniguchi}}, \bibinfo {author}
  {\bibfnamefont {D.~A.}\ \bibnamefont {Abanin}},\ and\ \bibinfo {author}
  {\bibfnamefont {A.~F.}\ \bibnamefont {Young}},\ }\href@noop {} {\bibfield
  {journal} {\bibinfo  {journal} {Nature Physics}\ }\textbf {\bibinfo {volume}
  {17}},\ \bibinfo {pages} {478} (\bibinfo {year}
  {2021}{\natexlab{b}})}\BibitemShut {NoStop}%
\bibitem [{\citenamefont {Andrei}\ and\ \citenamefont
  {MacDonald}(2021)}]{Andrei2021}%
  \BibitemOpen
  \bibfield  {author} {\bibinfo {author} {\bibfnamefont {E.~Y.}\ \bibnamefont
  {Andrei}}\ and\ \bibinfo {author} {\bibfnamefont {A.~H.}\ \bibnamefont
  {MacDonald}},\ }\href {https://doi.org/10.1038/s41563-020-00917-w} {\bibfield
   {journal} {\bibinfo  {journal} {Nature Materials}\ }\textbf {\bibinfo
  {volume} {20}},\ \bibinfo {pages} {571} (\bibinfo {year} {2021})}\BibitemShut
  {NoStop}%
\bibitem [{\citenamefont {Zhou}\ \emph
  {et~al.}(2021{\natexlab{a}})\citenamefont {Zhou}, \citenamefont {Xie},
  \citenamefont {Taniguchi}, \citenamefont {Watanabe},\ and\ \citenamefont
  {Young}}]{zhou2021superconductivity}%
  \BibitemOpen
  \bibfield  {author} {\bibinfo {author} {\bibfnamefont {H.}~\bibnamefont
  {Zhou}}, \bibinfo {author} {\bibfnamefont {T.}~\bibnamefont {Xie}}, \bibinfo
  {author} {\bibfnamefont {T.}~\bibnamefont {Taniguchi}}, \bibinfo {author}
  {\bibfnamefont {K.}~\bibnamefont {Watanabe}},\ and\ \bibinfo {author}
  {\bibfnamefont {A.~F.}\ \bibnamefont {Young}},\ }\href@noop {} {\bibfield
  {journal} {\bibinfo  {journal} {Nature}\ }\textbf {\bibinfo {volume} {598}},\
  \bibinfo {pages} {434} (\bibinfo {year} {2021}{\natexlab{a}})}\BibitemShut
  {NoStop}%
\bibitem [{\citenamefont {Zhou}\ \emph
  {et~al.}(2021{\natexlab{b}})\citenamefont {Zhou}, \citenamefont {Xie},
  \citenamefont {Ghazaryan}, \citenamefont {Holder}, \citenamefont {Ehrets},
  \citenamefont {Spanton}, \citenamefont {Taniguchi}, \citenamefont {Watanabe},
  \citenamefont {Berg}, \citenamefont {Serbyn} \emph {et~al.}}]{zhou2021half}%
  \BibitemOpen
  \bibfield  {author} {\bibinfo {author} {\bibfnamefont {H.}~\bibnamefont
  {Zhou}}, \bibinfo {author} {\bibfnamefont {T.}~\bibnamefont {Xie}}, \bibinfo
  {author} {\bibfnamefont {A.}~\bibnamefont {Ghazaryan}}, \bibinfo {author}
  {\bibfnamefont {T.}~\bibnamefont {Holder}}, \bibinfo {author} {\bibfnamefont
  {J.~R.}\ \bibnamefont {Ehrets}}, \bibinfo {author} {\bibfnamefont {E.~M.}\
  \bibnamefont {Spanton}}, \bibinfo {author} {\bibfnamefont {T.}~\bibnamefont
  {Taniguchi}}, \bibinfo {author} {\bibfnamefont {K.}~\bibnamefont {Watanabe}},
  \bibinfo {author} {\bibfnamefont {E.}~\bibnamefont {Berg}}, \bibinfo {author}
  {\bibfnamefont {M.}~\bibnamefont {Serbyn}}, \emph {et~al.},\ }\href@noop {}
  {\bibfield  {journal} {\bibinfo  {journal} {Nature}\ }\textbf {\bibinfo
  {volume} {598}},\ \bibinfo {pages} {429} (\bibinfo {year}
  {2021}{\natexlab{b}})}\BibitemShut {NoStop}%
\bibitem [{\citenamefont {Ghazaryan}\ \emph {et~al.}(2021)\citenamefont
  {Ghazaryan}, \citenamefont {Holder}, \citenamefont {Serbyn},\ and\
  \citenamefont {Berg}}]{ghazaryan2021unconventional}%
  \BibitemOpen
  \bibfield  {author} {\bibinfo {author} {\bibfnamefont {A.}~\bibnamefont
  {Ghazaryan}}, \bibinfo {author} {\bibfnamefont {T.}~\bibnamefont {Holder}},
  \bibinfo {author} {\bibfnamefont {M.}~\bibnamefont {Serbyn}},\ and\ \bibinfo
  {author} {\bibfnamefont {E.}~\bibnamefont {Berg}},\ }\href
  {https://doi.org/10.1103/PhysRevLett.127.247001} {\bibfield  {journal}
  {\bibinfo  {journal} {Phys. Rev. Lett.}\ }\textbf {\bibinfo {volume} {127}},\
  \bibinfo {pages} {247001} (\bibinfo {year} {2021})}\BibitemShut {NoStop}%
\bibitem [{\citenamefont {Chatterjee}\ \emph {et~al.}(2021)\citenamefont
  {Chatterjee}, \citenamefont {Wang}, \citenamefont {Berg},\ and\ \citenamefont
  {Zaletel}}]{chatterjee2021inter}%
  \BibitemOpen
  \bibfield  {author} {\bibinfo {author} {\bibfnamefont {S.}~\bibnamefont
  {Chatterjee}}, \bibinfo {author} {\bibfnamefont {T.}~\bibnamefont {Wang}},
  \bibinfo {author} {\bibfnamefont {E.}~\bibnamefont {Berg}},\ and\ \bibinfo
  {author} {\bibfnamefont {M.~P.}\ \bibnamefont {Zaletel}},\ }\href@noop {}
  {\bibfield  {journal} {\bibinfo  {journal} {arXiv preprint arXiv:2109.00002}\
  } (\bibinfo {year} {2021})}\BibitemShut {NoStop}%
\bibitem [{\citenamefont {You}\ and\ \citenamefont
  {Vishwanath}(2022)}]{You2022}%
  \BibitemOpen
  \bibfield  {author} {\bibinfo {author} {\bibfnamefont {Y.-Z.}\ \bibnamefont
  {You}}\ and\ \bibinfo {author} {\bibfnamefont {A.}~\bibnamefont
  {Vishwanath}},\ }\href {https://doi.org/10.1103/PhysRevB.105.134524}
  {\bibfield  {journal} {\bibinfo  {journal} {Phys. Rev. B}\ }\textbf {\bibinfo
  {volume} {105}},\ \bibinfo {pages} {134524} (\bibinfo {year}
  {2022})}\BibitemShut {NoStop}%
\bibitem [{\citenamefont {Chichinadze}\ \emph
  {et~al.}(2022{\natexlab{a}})\citenamefont {Chichinadze}, \citenamefont
  {Classen}, \citenamefont {Wang},\ and\ \citenamefont
  {Chubukov}}]{chichinadze_22}%
  \BibitemOpen
  \bibfield  {author} {\bibinfo {author} {\bibfnamefont {D.~V.}\ \bibnamefont
  {Chichinadze}}, \bibinfo {author} {\bibfnamefont {L.}~\bibnamefont
  {Classen}}, \bibinfo {author} {\bibfnamefont {Y.}~\bibnamefont {Wang}},\ and\
  \bibinfo {author} {\bibfnamefont {A.~V.}\ \bibnamefont {Chubukov}},\
  }\href@noop {} {\bibfield  {journal} {\bibinfo  {journal} {npj Quantum
  Materials}\ }\textbf {\bibinfo {volume} {7}},\ \bibinfo {pages} {114}
  (\bibinfo {year} {2022}{\natexlab{a}})}\BibitemShut {NoStop}%
\bibitem [{\citenamefont {McCann}\ and\ \citenamefont
  {Koshino}(2013)}]{McCann_2013}%
  \BibitemOpen
  \bibfield  {author} {\bibinfo {author} {\bibfnamefont {E.}~\bibnamefont
  {McCann}}\ and\ \bibinfo {author} {\bibfnamefont {M.}~\bibnamefont
  {Koshino}},\ }\href {https://doi.org/10.1088/0034-4885/76/5/056503}
  {\bibfield  {journal} {\bibinfo  {journal} {Reports on Progress in Physics}\
  }\textbf {\bibinfo {volume} {76}},\ \bibinfo {pages} {056503} (\bibinfo
  {year} {2013})}\BibitemShut {NoStop}%
\bibitem [{\citenamefont {McCann}\ and\ \citenamefont
  {Fal'ko}(2006)}]{McCann2006Landau}%
  \BibitemOpen
  \bibfield  {author} {\bibinfo {author} {\bibfnamefont {E.}~\bibnamefont
  {McCann}}\ and\ \bibinfo {author} {\bibfnamefont {V.~I.}\ \bibnamefont
  {Fal'ko}},\ }\href {https://doi.org/10.1103/PhysRevLett.96.086805} {\bibfield
   {journal} {\bibinfo  {journal} {Phys. Rev. Lett.}\ }\textbf {\bibinfo
  {volume} {96}},\ \bibinfo {pages} {086805} (\bibinfo {year}
  {2006})}\BibitemShut {NoStop}%
\bibitem [{\citenamefont {McCann}(2006)}]{McCann2006Asymmetry}%
  \BibitemOpen
  \bibfield  {author} {\bibinfo {author} {\bibfnamefont {E.}~\bibnamefont
  {McCann}},\ }\href {https://doi.org/10.1103/PhysRevB.74.161403} {\bibfield
  {journal} {\bibinfo  {journal} {Phys. Rev. B}\ }\textbf {\bibinfo {volume}
  {74}},\ \bibinfo {pages} {161403} (\bibinfo {year} {2006})}\BibitemShut
  {NoStop}%
\bibitem [{\citenamefont {Nandkishore}\ and\ \citenamefont
  {Levitov}(2010{\natexlab{a}})}]{Nandkishore2010dynamical}%
  \BibitemOpen
  \bibfield  {author} {\bibinfo {author} {\bibfnamefont {R.}~\bibnamefont
  {Nandkishore}}\ and\ \bibinfo {author} {\bibfnamefont {L.}~\bibnamefont
  {Levitov}},\ }\href {https://doi.org/10.1103/PhysRevLett.104.156803}
  {\bibfield  {journal} {\bibinfo  {journal} {Phys. Rev. Lett.}\ }\textbf
  {\bibinfo {volume} {104}},\ \bibinfo {pages} {156803} (\bibinfo {year}
  {2010}{\natexlab{a}})}\BibitemShut {NoStop}%
\bibitem [{\citenamefont {Nandkishore}\ and\ \citenamefont
  {Levitov}(2010{\natexlab{b}})}]{Nandkishore2010quantum}%
  \BibitemOpen
  \bibfield  {author} {\bibinfo {author} {\bibfnamefont {R.}~\bibnamefont
  {Nandkishore}}\ and\ \bibinfo {author} {\bibfnamefont {L.}~\bibnamefont
  {Levitov}},\ }\href {https://doi.org/10.1103/PhysRevB.82.115124} {\bibfield
  {journal} {\bibinfo  {journal} {Phys. Rev. B}\ }\textbf {\bibinfo {volume}
  {82}},\ \bibinfo {pages} {115124} (\bibinfo {year}
  {2010}{\natexlab{b}})}\BibitemShut {NoStop}%
\bibitem [{\citenamefont {Vafek}\ and\ \citenamefont {Yang}(2010)}]{Vafek2010}%
  \BibitemOpen
  \bibfield  {author} {\bibinfo {author} {\bibfnamefont {O.}~\bibnamefont
  {Vafek}}\ and\ \bibinfo {author} {\bibfnamefont {K.}~\bibnamefont {Yang}},\
  }\href {https://doi.org/10.1103/PhysRevB.81.041401} {\bibfield  {journal}
  {\bibinfo  {journal} {Phys. Rev. B}\ }\textbf {\bibinfo {volume} {81}},\
  \bibinfo {pages} {041401} (\bibinfo {year} {2010})}\BibitemShut {NoStop}%
\bibitem [{\citenamefont {Jung}\ \emph {et~al.}(2011)\citenamefont {Jung},
  \citenamefont {Zhang},\ and\ \citenamefont {MacDonald}}]{Jung2011lattice}%
  \BibitemOpen
  \bibfield  {author} {\bibinfo {author} {\bibfnamefont {J.}~\bibnamefont
  {Jung}}, \bibinfo {author} {\bibfnamefont {F.}~\bibnamefont {Zhang}},\ and\
  \bibinfo {author} {\bibfnamefont {A.~H.}\ \bibnamefont {MacDonald}},\ }\href
  {https://doi.org/10.1103/PhysRevB.83.115408} {\bibfield  {journal} {\bibinfo
  {journal} {Phys. Rev. B}\ }\textbf {\bibinfo {volume} {83}},\ \bibinfo
  {pages} {115408} (\bibinfo {year} {2011})}\BibitemShut {NoStop}%
\bibitem [{\citenamefont {MacDonald}\ \emph {et~al.}(2012)\citenamefont
  {MacDonald}, \citenamefont {Jung},\ and\ \citenamefont
  {Zhang}}]{macdonald2012pseudospin}%
  \BibitemOpen
  \bibfield  {author} {\bibinfo {author} {\bibfnamefont {A.~H.}\ \bibnamefont
  {MacDonald}}, \bibinfo {author} {\bibfnamefont {J.}~\bibnamefont {Jung}},\
  and\ \bibinfo {author} {\bibfnamefont {F.}~\bibnamefont {Zhang}},\
  }\href@noop {} {\bibfield  {journal} {\bibinfo  {journal} {Physica Scripta}\
  }\textbf {\bibinfo {volume} {2012}},\ \bibinfo {pages} {014012} (\bibinfo
  {year} {2012})}\BibitemShut {NoStop}%
\bibitem [{\citenamefont {Zhang}\ and\ \citenamefont
  {MacDonald}(2012)}]{Zhang2012Distinguishing}%
  \BibitemOpen
  \bibfield  {author} {\bibinfo {author} {\bibfnamefont {F.}~\bibnamefont
  {Zhang}}\ and\ \bibinfo {author} {\bibfnamefont {A.~H.}\ \bibnamefont
  {MacDonald}},\ }\href {https://doi.org/10.1103/PhysRevLett.108.186804}
  {\bibfield  {journal} {\bibinfo  {journal} {Phys. Rev. Lett.}\ }\textbf
  {\bibinfo {volume} {108}},\ \bibinfo {pages} {186804} (\bibinfo {year}
  {2012})}\BibitemShut {NoStop}%
\bibitem [{\citenamefont {Cvetkovic}\ \emph {et~al.}(2012)\citenamefont
  {Cvetkovic}, \citenamefont {Throckmorton},\ and\ \citenamefont
  {Vafek}}]{Cvetkovic2012}%
  \BibitemOpen
  \bibfield  {author} {\bibinfo {author} {\bibfnamefont {V.}~\bibnamefont
  {Cvetkovic}}, \bibinfo {author} {\bibfnamefont {R.~E.}\ \bibnamefont
  {Throckmorton}},\ and\ \bibinfo {author} {\bibfnamefont {O.}~\bibnamefont
  {Vafek}},\ }\href {https://doi.org/10.1103/PhysRevB.86.075467} {\bibfield
  {journal} {\bibinfo  {journal} {Phys. Rev. B}\ }\textbf {\bibinfo {volume}
  {86}},\ \bibinfo {pages} {075467} (\bibinfo {year} {2012})}\BibitemShut
  {NoStop}%
\bibitem [{\citenamefont {Throckmorton}\ and\ \citenamefont
  {Das~Sarma}(2014)}]{Throckmorton2014}%
  \BibitemOpen
  \bibfield  {author} {\bibinfo {author} {\bibfnamefont {R.~E.}\ \bibnamefont
  {Throckmorton}}\ and\ \bibinfo {author} {\bibfnamefont {S.}~\bibnamefont
  {Das~Sarma}},\ }\href {https://doi.org/10.1103/PhysRevB.90.205407} {\bibfield
   {journal} {\bibinfo  {journal} {Phys. Rev. B}\ }\textbf {\bibinfo {volume}
  {90}},\ \bibinfo {pages} {205407} (\bibinfo {year} {2014})}\BibitemShut
  {NoStop}%
\bibitem [{\citenamefont {Min}\ \emph {et~al.}(2008)\citenamefont {Min},
  \citenamefont {Borghi}, \citenamefont {Polini},\ and\ \citenamefont
  {MacDonald}}]{Min2008}%
  \BibitemOpen
  \bibfield  {author} {\bibinfo {author} {\bibfnamefont {H.}~\bibnamefont
  {Min}}, \bibinfo {author} {\bibfnamefont {G.}~\bibnamefont {Borghi}},
  \bibinfo {author} {\bibfnamefont {M.}~\bibnamefont {Polini}},\ and\ \bibinfo
  {author} {\bibfnamefont {A.~H.}\ \bibnamefont {MacDonald}},\ }\href
  {https://doi.org/10.1103/PhysRevB.77.041407} {\bibfield  {journal} {\bibinfo
  {journal} {Phys. Rev. B}\ }\textbf {\bibinfo {volume} {77}},\ \bibinfo
  {pages} {041407} (\bibinfo {year} {2008})}\BibitemShut {NoStop}%
\bibitem [{\citenamefont {Nilsson}\ \emph {et~al.}(2006)\citenamefont
  {Nilsson}, \citenamefont {Castro~Neto}, \citenamefont {Peres},\ and\
  \citenamefont {Guinea}}]{Nilsson2006}%
  \BibitemOpen
  \bibfield  {author} {\bibinfo {author} {\bibfnamefont {J.}~\bibnamefont
  {Nilsson}}, \bibinfo {author} {\bibfnamefont {A.~H.}\ \bibnamefont
  {Castro~Neto}}, \bibinfo {author} {\bibfnamefont {N.~M.~R.}\ \bibnamefont
  {Peres}},\ and\ \bibinfo {author} {\bibfnamefont {F.}~\bibnamefont
  {Guinea}},\ }\href {https://doi.org/10.1103/PhysRevB.73.214418} {\bibfield
  {journal} {\bibinfo  {journal} {Phys. Rev. B}\ }\textbf {\bibinfo {volume}
  {73}},\ \bibinfo {pages} {214418} (\bibinfo {year} {2006})}\BibitemShut
  {NoStop}%
\bibitem [{\citenamefont {Pantaleon}\ \emph {et~al.}(2022)\citenamefont
  {Pantaleon}, \citenamefont {Jimeno-Pozo}, \citenamefont {Sainz-Cruz},
  \citenamefont {Cea}, \citenamefont {Phong},\ and\ \citenamefont
  {Guinea}}]{pantaleon2022superconductivity}%
  \BibitemOpen
  \bibfield  {author} {\bibinfo {author} {\bibfnamefont {P.~A.}\ \bibnamefont
  {Pantaleon}}, \bibinfo {author} {\bibfnamefont {A.}~\bibnamefont
  {Jimeno-Pozo}}, \bibinfo {author} {\bibfnamefont {H.}~\bibnamefont
  {Sainz-Cruz}}, \bibinfo {author} {\bibfnamefont {T.}~\bibnamefont {Cea}},
  \bibinfo {author} {\bibfnamefont {V.~T.}\ \bibnamefont {Phong}},\ and\
  \bibinfo {author} {\bibfnamefont {F.}~\bibnamefont {Guinea}},\ }\href@noop {}
  {\bibfield  {journal} {\bibinfo  {journal} {arXiv preprint arXiv:2211.02880}\
  } (\bibinfo {year} {2022})}\BibitemShut {NoStop}%
\bibitem [{\citenamefont {Dong}\ \emph {et~al.}(2021)\citenamefont {Dong},
  \citenamefont {Davydova}, \citenamefont {Ogunnaike},\ and\ \citenamefont
  {Levitov}}]{dong2021isospin}%
  \BibitemOpen
  \bibfield  {author} {\bibinfo {author} {\bibfnamefont {Z.}~\bibnamefont
  {Dong}}, \bibinfo {author} {\bibfnamefont {M.}~\bibnamefont {Davydova}},
  \bibinfo {author} {\bibfnamefont {O.}~\bibnamefont {Ogunnaike}},\ and\
  \bibinfo {author} {\bibfnamefont {L.}~\bibnamefont {Levitov}},\ }\href@noop
  {} {\bibfield  {journal} {\bibinfo  {journal} {arXiv preprint
  arXiv:2110.15254}\ } (\bibinfo {year} {2021})}\BibitemShut {NoStop}%
\bibitem [{\citenamefont {Chichinadze}\ \emph
  {et~al.}(2022{\natexlab{b}})\citenamefont {Chichinadze}, \citenamefont
  {Classen}, \citenamefont {Wang},\ and\ \citenamefont
  {Chubukov}}]{chichinadze_prl}%
  \BibitemOpen
  \bibfield  {author} {\bibinfo {author} {\bibfnamefont {D.~V.}\ \bibnamefont
  {Chichinadze}}, \bibinfo {author} {\bibfnamefont {L.}~\bibnamefont
  {Classen}}, \bibinfo {author} {\bibfnamefont {Y.}~\bibnamefont {Wang}},\ and\
  \bibinfo {author} {\bibfnamefont {A.~V.}\ \bibnamefont {Chubukov}},\ }\href
  {https://doi.org/10.1103/PhysRevLett.128.227601} {\bibfield  {journal}
  {\bibinfo  {journal} {Phys. Rev. Lett.}\ }\textbf {\bibinfo {volume} {128}},\
  \bibinfo {pages} {227601} (\bibinfo {year} {2022}{\natexlab{b}})}\BibitemShut
  {NoStop}%
\bibitem [{\citenamefont {Dong}\ \emph {et~al.}(2022)\citenamefont {Dong},
  \citenamefont {Chubukov},\ and\ \citenamefont {Levitov}}]{dong_23}%
  \BibitemOpen
  \bibfield  {author} {\bibinfo {author} {\bibfnamefont {Z.}~\bibnamefont
  {Dong}}, \bibinfo {author} {\bibfnamefont {A.~V.}\ \bibnamefont {Chubukov}},\
  and\ \bibinfo {author} {\bibfnamefont {L.}~\bibnamefont {Levitov}},\ }\href
  {https://doi.org/10.48550/ARXIV.2205.13353} {\bibinfo {title} {Spin-triplet
  superconductivity at the onset of isospin order in biased bilayer graphene}}
  (\bibinfo {year} {2022})\BibitemShut {NoStop}%
\bibitem [{\citenamefont {Zhou}\ \emph
  {et~al.}(2021{\natexlab{c}})\citenamefont {Zhou}, \citenamefont {Saito},
  \citenamefont {Cohen}, \citenamefont {Huynh}, \citenamefont {Patterson},
  \citenamefont {Yang}, \citenamefont {Taniguchi}, \citenamefont {Watanabe},\
  and\ \citenamefont {Young}}]{zhou2021BLG}%
  \BibitemOpen
  \bibfield  {author} {\bibinfo {author} {\bibfnamefont {H.}~\bibnamefont
  {Zhou}}, \bibinfo {author} {\bibfnamefont {Y.}~\bibnamefont {Saito}},
  \bibinfo {author} {\bibfnamefont {L.}~\bibnamefont {Cohen}}, \bibinfo
  {author} {\bibfnamefont {W.}~\bibnamefont {Huynh}}, \bibinfo {author}
  {\bibfnamefont {C.~L.}\ \bibnamefont {Patterson}}, \bibinfo {author}
  {\bibfnamefont {F.}~\bibnamefont {Yang}}, \bibinfo {author} {\bibfnamefont
  {T.}~\bibnamefont {Taniguchi}}, \bibinfo {author} {\bibfnamefont
  {K.}~\bibnamefont {Watanabe}},\ and\ \bibinfo {author} {\bibfnamefont
  {A.~F.}\ \bibnamefont {Young}},\ }\href@noop {} {\bibfield  {journal}
  {\bibinfo  {journal} {arXiv preprint arXiv:2110.11317}\ } (\bibinfo {year}
  {2021}{\natexlab{c}})}\BibitemShut {NoStop}%
\bibitem [{\citenamefont {Scalapino}(2012)}]{Scalapino2012}%
  \BibitemOpen
  \bibfield  {author} {\bibinfo {author} {\bibfnamefont {D.~J.}\ \bibnamefont
  {Scalapino}},\ }\href {https://doi.org/10.1103/RevModPhys.84.1383} {\bibfield
   {journal} {\bibinfo  {journal} {Rev. Mod. Phys.}\ }\textbf {\bibinfo
  {volume} {84}},\ \bibinfo {pages} {1383} (\bibinfo {year}
  {2012})}\BibitemShut {NoStop}%
\bibitem [{\citenamefont {Monthoux}\ \emph {et~al.}(2007)\citenamefont
  {Monthoux}, \citenamefont {Pines},\ and\ \citenamefont
  {Lonzarich}}]{Monthoux2007}%
  \BibitemOpen
  \bibfield  {author} {\bibinfo {author} {\bibfnamefont {P.}~\bibnamefont
  {Monthoux}}, \bibinfo {author} {\bibfnamefont {D.}~\bibnamefont {Pines}},\
  and\ \bibinfo {author} {\bibfnamefont {G.~G.}\ \bibnamefont {Lonzarich}},\
  }\href {https://doi.org/10.1038/nature06480} {\bibfield  {journal} {\bibinfo
  {journal} {Nature}\ }\textbf {\bibinfo {volume} {450}},\ \bibinfo {pages}
  {1177} (\bibinfo {year} {2007})}\BibitemShut {NoStop}%
\bibitem [{\citenamefont {Chichinadze}\ \emph {et~al.}(2020)\citenamefont
  {Chichinadze}, \citenamefont {Classen},\ and\ \citenamefont
  {Chubukov}}]{chichinadze_20}%
  \BibitemOpen
  \bibfield  {author} {\bibinfo {author} {\bibfnamefont {D.~V.}\ \bibnamefont
  {Chichinadze}}, \bibinfo {author} {\bibfnamefont {L.}~\bibnamefont
  {Classen}},\ and\ \bibinfo {author} {\bibfnamefont {A.~V.}\ \bibnamefont
  {Chubukov}},\ }\href {https://doi.org/10.1103/PhysRevB.102.125120} {\bibfield
   {journal} {\bibinfo  {journal} {Phys. Rev. B}\ }\textbf {\bibinfo {volume}
  {102}},\ \bibinfo {pages} {125120} (\bibinfo {year} {2020})}\BibitemShut
  {NoStop}%
\bibitem [{\citenamefont {Chichinadze}\ \emph {et~al.}(2021)\citenamefont
  {Chichinadze}, \citenamefont {Classen},\ and\ \citenamefont
  {Chubukov}}]{chichinadze_20a}%
  \BibitemOpen
  \bibfield  {author} {\bibinfo {author} {\bibfnamefont {D.~V.}\ \bibnamefont
  {Chichinadze}}, \bibinfo {author} {\bibfnamefont {L.}~\bibnamefont
  {Classen}},\ and\ \bibinfo {author} {\bibfnamefont {A.~V.}\ \bibnamefont
  {Chubukov}},\ }\href {https://doi.org/10.1103/PhysRevB.103.039901} {\bibfield
   {journal} {\bibinfo  {journal} {Phys. Rev. B}\ }\textbf {\bibinfo {volume}
  {103}},\ \bibinfo {pages} {039901} (\bibinfo {year} {2021})}\BibitemShut
  {NoStop}%
\bibitem [{\citenamefont {You}\ and\ \citenamefont
  {Vishwanath}(2021)}]{You_21}%
  \BibitemOpen
  \bibfield  {author} {\bibinfo {author} {\bibfnamefont {Y.-Z.}\ \bibnamefont
  {You}}\ and\ \bibinfo {author} {\bibfnamefont {A.}~\bibnamefont
  {Vishwanath}},\ }\href {https://doi.org/10.48550/ARXIV.2109.04669} {\bibinfo
  {title} {Kohn-luttinger superconductivity and inter-valley coherence in
  rhombohedral trilayer graphene}} (\bibinfo {year} {2021})\BibitemShut
  {NoStop}%
\bibitem [{\citenamefont {Chatterjee}\ \emph {et~al.}(2022)\citenamefont
  {Chatterjee}, \citenamefont {Wang}, \citenamefont {Berg},\ and\ \citenamefont
  {Zaletel}}]{chatterjee2022inter}%
  \BibitemOpen
  \bibfield  {author} {\bibinfo {author} {\bibfnamefont {S.}~\bibnamefont
  {Chatterjee}}, \bibinfo {author} {\bibfnamefont {T.}~\bibnamefont {Wang}},
  \bibinfo {author} {\bibfnamefont {E.}~\bibnamefont {Berg}},\ and\ \bibinfo
  {author} {\bibfnamefont {M.~P.}\ \bibnamefont {Zaletel}},\ }\href@noop {}
  {\bibfield  {journal} {\bibinfo  {journal} {Nature Communications}\ }\textbf
  {\bibinfo {volume} {13}},\ \bibinfo {pages} {6013} (\bibinfo {year}
  {2022})}\BibitemShut {NoStop}%
\end{thebibliography}%

	\end{document}